\documentclass[12pt, a4paper]{article}

\pdfoutput=1

\usepackage{amsmath}
\usepackage{amsfonts}
\usepackage{amssymb}
\usepackage{latexsym}
\usepackage{graphicx}
\usepackage{color}
\usepackage{mathrsfs}
\usepackage{slashed}
\usepackage{soul}
\usepackage{array}
\usepackage{cite}
\usepackage{hyperref}
\usepackage{enumerate}
\usepackage{multirow}
\usepackage[table]{xcolor}
\usepackage{colortbl}
\usepackage{float}
\usepackage{comment}
\includecomment{toexclude} 
\excludecomment{addsections} 

\def\ga   {\gamma}

\def\th   {\theta}
\def\la   {\lambda}
\def\al   {\alpha}
\def\sig   {\sigma}

\def\nn{\nonumber}
\def\lee { \left( }
\def\rii { \right) }
\def\lan   {\langle}
\def\ran   {\rangle}

\def\tca {\text{Hyb}\nu_\tau}
\def\mca {\text{Hyb}\nu_\mu}
\def\eca {\text{Hyb}\nu_e}

\def\cfot {\cos^2\theta_{41}}
\def\sfot {\sin^2\theta_{41}}

\begin{document}

\vspace*{0.5cm}

\begin{center}
{\Huge
Dark Matter and Lepton Flavour Violation\\[2mm]
in a Hybrid Neutrino Mass Model
}
\\ [2.5cm]
{\large{
Frank Deppisch\footnote{\textsl{f.deppisch@ucl.ac.uk}},
Wei-Chih Huang\footnote{\textsl{wei-chih.huang@ucl.ac.uk}} }}
\\[0.5cm]

\large{\textit{
Department of Physics and Astronomy, University College London, UK
}}
\\ [2cm]
{ \large{\textrm{
Abstract
}}}
\\ [0.5cm]
\end{center}
We describe a hybrid model in which the light neutrino mass matrix receives both tree-level seesaw and loop-induced contributions. An additional $U(1)$ gauge symmetry is used to stabilize the lightest right-handed neutrino as the Dark Matter candidate. After fitting the experimental neutrino data, we analyze and correlate the phenomenological consequences of the model, namely its impact on electroweak precision measurements, the Dark Matter relic abundance, lepton flavour violating rare decays and neutrinoless double beta decay. We find that natural realizations of the model characterized by large Yukawa couplings are compatible with and close to the current experimental limits.

\def\thefootnote{\arabic{footnote}}
\setcounter{footnote}{0}
\pagestyle{empty}

\newpage
\pagestyle{plain}

\section{Introduction}
\label{sec:intro}

The observation of neutrino oscillations~\cite{fukuda:1998mi, ahmad:2002jz, eguchi:2002dm} has essentially established that at least two of the active neutrinos have a tiny yet finite mass~\cite{Maltoni:2004ei} and that individual lepton flavours are not conserved in nature. Especially after the discovery of the Higgs boson which strongly indicates a mechanism that provides masses to the charged fermions, the question of why neutrinos have a small but finite mass is one of the key issues in particle physics.

There are many new physics mechanisms that provide neutrino masses in the required regime. Certainly the most popular is the so-called seesaw mechanism where sterile right-handed Majorana neutrinos are added to the Standard Model (SM) particle spectrum \cite{minkowski:1977sc, gell-mann:1980vs, yanagida:1979, mohapatra:1979ia, schechter:1980gr, schechter:1981cv, lazarides:1980nt}. Their mixing with the left-handed neutrinos, driven by electroweak (EW) symmetry breaking, then generates the tiny masses of the active neutrinos. This seesaw type-I mechanism constitutes one of three ways to ultraviolet-complete the five-dimensional Weinberg operator at tree level.

Although very suggestive and theoretically well motivated, the default seesaw type-I mechanism has major phenomenological problems: super-heavy neutrinos with masses of the order $10^{14}$~GeV cannot be probed directly, and they are gauge singlets. The latter means that even if they were light enough to be accessible in experiments, they only couple to SM particles with their small Yukawa interactions. The simplest realizations of the seesaw mechanism are therefore difficult to probe in experiments such as the LHC~\cite{delAguila:2007em}. Extended realizations of the seesaw idea such as inverse seesaw~\cite{Mohapatra:1986bd}, with right-handed neutrino masses of the order of or close to the EW scale, but with an approximate conservation of lepton number provide a way out of this problem. The right-handed neutrinos are still gauge singlets (unless the gauge symmetry is expanded appropriately as in left-right symmetric models), but their Yukawa couplings can be large.

In seesaw models, light neutrino masses are generated by breaking lepton number at a high scale and thereby inducing a small effective coupling of the five-dimensional Weinberg operator at tree level. Alternatively, this smallness may also be understood by inducing the operator radiatively, i.e. at one or higher loop order \cite{Zee:1980ai, Babu:1988ki, Pilaftsis:1991ug, Dev:2012sg}. Symmetries employed to forbid the tree level mediation of the breaking of lepton number to the visible sector can also be used to provide a stable Dark Matter (DM) candidate \cite{Krauss:2002px, Cheung:2004xm, Ma:2006km, Aoki:2008av, Aoki:2009vf}.  

In this work, we describe a hybrid model that incorporates both a tree level  (inverse) seesaw type-I and a loop-induced contribution to the light neutrino mass matrix. The model itself is based on the ``scotogenic'' model of Ref.~\cite{Ma:2006km}, which provides the simplest realization for a radiative neutrino mass, where an additional $SU(2)_L$ scalar doublet and the heavy right-handed neutrinos are charged under a $Z_2$ symmetry. The lightest of the $Z_2$-odd particles is stable and can be a DM candidate. In this work, we concentrate on the case of the lightest right-handed $N_1$ being the cold DM particle~\cite{Kubo:2006yx}. There, however, exists some tension between having the correct relic density and the stringent lepton flavour violation~(LFV) $\mu \to e \gamma$ bounds, although some regions of the parameter space are still allowed~\cite{Toma:2013zsa}. It could be circumvented with the extension of the fermion content or involving co-annihilations~\cite{Ma:2008cu, Suematsu:2009ww, Schmidt:2012yg,Hirsch:2013ola,Brdar:2013iea} or to include additional interactions for $N_1$~\cite{Kubo:2006rm, Suematsu:2007dc, Babu:2007sm,Ma:2013yga}. 

Our model, with an additional $U(1)_X$ gauge group, has a few distinctive features. First, it is a hybrid model consisting of both the seesaw type-I and radiative contribution to the neutrino mass. Second, except for the Higgs boson and one generation of leptons, all fermions are charged under the $U(1)_X$ gauge group. Third, the mass difference between the neutral components of the additional $SU(2)_L$ scalar doublet is naturally small. The small neutrino masses are therefore effectively induced at second loop order. Finally, the loop-induced LFV process $\mu \to e\gamma$ can be suppressed through a GIM-like mechanism when the heavy neutrino masses are degenerate, whereas another transition like $\tau \to \mu\gamma$ is generated by  the light-heavy neutrino mixing as well as the loop correction, and can be large and unsuppressed for quasi-degenerate heavy neutrinos.

Our paper is organized as follows. In Section~\ref{sec:framework}, we present our model under consideration, in turn addressing the symmetry structure and the particle spectrum. We then proceed by discussing the most relevant observables, namely the neutrino masses and mixing, charged lepton flavour violating decays and the relic abundance in Section~\ref{fitting_nu_data}, \ref{sec:LFV} and \ref{sec:DM}, respectively. Combining the experimental constraints on these observables, we numerically analyze the parameter space of the model in Section~\ref{sec:result}. Finally, we summarize our findings and address further aspects of the model in Section~\ref{sec:conclusions}.

\section{Model Framework}\label{sec:framework}

\subsection{Model description}\label{sec:model}
Our model extends the SM by introducing an additional $U(1)_X$ gauge group. In addition to the SM particle content, we introduce at least four sterile, right-handed neutrinos $N_i$ (as discussed later, at least four of them are needed in order to make the theory anomaly-free) as well as an exotic $SU(2)_L$ scalar doublet $\phi$ and the $U(1)_X$ gauge boson $Z^\prime$.
The SM fermions, the $N_i$ and the $\phi$ are all charged under the $U(1)_X$. To accommodate a stable DM particle, we resort to a $U(1)_X$ charge assignment such that $N_{1,4}$  and $\phi$ have odd $U(1)_X$ charges and the rest of the fields carry even charges. This ensures the stability of the lightest oddly charged particle, which we take to be the lightest heavy neutrino $N_1$. Although not directly necessary for our discussion we assume that the $U(1)_X$ gauge symmetry is broken spontaneously, providing Majorana masses to the $N_i$. This can, for example, be achieved by introducing scalars $\Phi_i$, charged under  $U(1)_X$ such that $\lan \Phi \ran N N$ yields the Majorana mass term and at the same time $Z^\prime$ receives a mass $m_{Z'} \approx g_X \lan \Phi \ran$. Note that we neglect the mixing between $U(1)_X$ and the SM $U(1)_Y$, which is generally small since SM fermions are also charged
under $U(1)_X$.

In the following, we summarize the model quantum numbers and present the full Lagrangian, followed by detailed discussions of the particle spectrum in the succeeding sections. The chosen quantum numbers of the relevant particles are shown in Tab.~\ref{tab:quantum_number}, based on the anomaly cancellation consideration as discussed below. The lepton doublets are denoted by $L_\alpha \equiv (\nu_\alpha \,\, e_\alpha)^T$ with $\alpha=(e, \mu, \tau)$ and $H$ is the SM Higgs doublet, which is neutral under $U(1)_X$. Two of three lepton generations, $L_\alpha$ and $L_\beta$ carry a $U(1)_X$ charge of two while one generation $L_\gamma$ is neutral under $U(1)_X$. This choice enables the hybrid generation of neutrino masses such that $\nu_{\alpha, \beta}$ receive a radiative mass from their interaction with $N_1$ and $\nu_{\gamma}$ obtains one from $N_4$. On the other hand, only $\nu_{\alpha,\beta}$ have Yukawa couplings with the heavy neutrinos $N_{2,3}$ and acquire a tree level mass via the seesaw type-I mechanism. In the following, we refer to the three possible cases of assigning charges by $\eca$, $\mca$ and $\tca$ for $\gamma=e$, $\gamma=\mu$ and $\gamma=\tau$, respectively. Please note that these are not the only possible choices and other scenarios can be considered, but for clarity we focus on the above three cases to illustrate the flavour structure in the hybrid neutrino mass model discussed here.
\begin{table}[t!]
\centering
\begin{tabular}{cccccccccc c c c}
\hline
Group & $L_{\alpha,\beta}$ & $L_{\gamma}$ & $E_{\alpha,\beta}$ &  $E_{\ga}$ &$H$ & $N_1$ & $N_4$ &$N_{2,3}$ & $\phi$
& Q & U & D
\\\hline
$SU(2)_L$ &   2   & 2   & 1 & 1& 2   &  1 & 1 &  1 &   2   & 2 & 1 & 1
\\
$U(1)_Y$  &  -1/2 & -1/2 & 1 & 1 & 1/2 &  0 & 0 &  0 &   1/2 & 1/6 & -2/3 & 1/3
\\
$U(1)_X$  &   2   & 0  & -2  & 0 & 0   & -1 & 1 & -2 &  -1   & -4/9 & 4/9 & 4/9
\\
\hline
\end{tabular}
\caption{Quantum numbers of exotic and relevant SM fields under the SM gauge groups $SU(2)_L$, $U(1)_Y$ and the extra $U(1)_X $.
The lepton~(quark) doublet is denoted by $L$~(Q) while $E$, $U$ and $D$ refer to the $SU(2)_L$ singlet charge lepton, up-type quark and down-type quark,
respectively.
The three lepton flavours are denoted by the generic indices $\alpha$, $\beta$ and $\gamma$, i.e. two of three lepton doublets carry $X_L=2$ and the remaining has $X_L$=0. }
\label{tab:quantum_number}
\end{table}

In this paper, we use the two component Weyl-spinor notation and fields without~(with) ``$\, \dag \,$'' are left-handed~(right-handed), e.g. $N_1$ represents the left-handed field of the lightest heavy neutrino while $N_1^\dag$ corresponds to the right-handed field. Therefore, the $U(1)_X$ charge of $N_1^\dag$ is opposite to that of $N_1$, i.e., $X_{N_1^\dag}= - X_{N_1}$. Suppressing kinetic terms, the Lagrangian for the model reads 
\begin{align}
	\mathcal{L} \supset 
	\mathcal{L}_{SM}+ \mathcal{L}_{y} + \mathcal{L}_N + 	\mathcal{L}_{\phi^2},
\end{align} 
where $\mathcal{L}_{SM}$ is the SM Lagrangian and 
\begin{align}
	\mathcal{L}_y &= \sum^3_{i=2} \sum_{\rho=\alpha,\beta} 
	y_{\rho i } \lee L_{\rho} \cdot H \rii N_{i}
	+ \sum_{\rho=\alpha,\beta} \la_\rho \lee L_{\rho} \cdot \phi \rii N_1
	+ \la_{N_4} \lee L_{\ga} \cdot \phi \rii N_4 + h.c.  \, , \label{eq:Ly} \\
	\mathcal{L}_N &= 
	  \frac{1}{2} \sum_{i,j=2,3} (m_N)_{ij} N_i N_j 
	+ \frac{1}{2} \sum_{i,j=1,4} (m_N)_{ij} N_i N_j 
	+ h.c., \label{eq:LN}\\
	\mathcal{L}_{\phi^2} &= m_\phi^{\prime 2} \phi^\dag \phi
	+ c_1 (\tilde{H} \cdot H) (\phi^\dag \cdot\phi) 
	+ c_2 (\tilde{H} \cdot \phi) (H  \cdot\phi^\dag)
	+ \lee c_3 (\tilde{H} \cdot \phi) (\tilde{H} \cdot \phi)  
	+ h.c. \rii. 
	\label{eq:Lphi}
\end{align}
Here, the dot denotes the $SU(2)$-invariant product of doublets. The first part
$\mathcal{L}_y$ corresponds to Yukawa-type interactions of the left-handed doublets with the heavy neutrinos, leading to both the seesaw type-I and radiative mass contributions to the light neutrino masses. The second component $\mathcal{L}_N$ describes the heavy neutrino mass terms, which we assume are generated by breaking the $U(1)_X$ in a hidden sector such that $N_2$, $N_3$ and $N_1$, $N_4$ may mix but not between the two species. Finally, $\mathcal{L}_{\phi^2}$ consists of mass terms of $\phi$ that are relevant to the generation of radiative masses of the light neutrinos.  

Anomaly cancellation among the SM gauge groups, the $U(1)_X$ and gravity~($G$)~\cite{Carena:2004xs} puts constraints on the $U(1)_X$ charges of particles of interest. The cancellation conditions are presented in detail in Appendix~\ref{sec:anomaly} and we summarize the results here.
The fourth heavy neutrino $N_4$ on top of $N_{1,2,3}$ is needed, otherwise one of the lepton doublets will carry the same $U(1)_X$ charge as $N_1$ in order to make the model anomaly-free; it renders $N_1$ unstable, decaying into the Higgs boson and light neutrino.  As a consequence, $N_1$, $N_4$ and $\phi$ are assigned odd charges, with others carrying even charges such that $N_1$ can be a stable DM candidate. Furthermore, we stick to integer charges for simplicity\footnote{For the recent study involving irrational charges in the context of $U(1)_{B-L}$, see for example Ref.~\cite{Kanemura:2014rpa}.}. There are three possible charge assignments:
\begin{itemize}
\item $X_{N_4}= - X_{N_1}=1$, $X_{N_2}=X_{N_3}=-2$, $X_{L_\mu}=X_{L_\tau}=2$, $X_{L_e}=0$,
\item $X_{N_4}= - X_{N_1}=1$, $X_{N_2}=X_{N_3}=-2$, $X_{L_e}=X_{L_\tau}=2$, $X_{L_\mu}=0$,
\item $X_{N_4}= - X_{N_1}=1$, $X_{N_2}=X_{N_3}=-2$, $X_{L_e}=X_{L_\mu}=2$, $X_{L_\tau}=0$,
\end{itemize}
We refer to these cases as $\eca$, $\mca$ and $\tca$, respectively. Each of them corresponds to a different flavour structure of the light neutrino mass matrix. For example, in the case $\eca$, the block $\nu_\mu$-$\nu_\tau$ receives both seesaw type-I and $N_1$-radiative contributions, while loops involving $N_4$ give a mass to $\nu_e$ only. Besides, the $N_1-N_4$ mixing induces radiative mass terms of $\nu_e-\nu_{\mu,\tau}$.

\subsection{Particle Spectrum}
\label{sec:particle spectrum}

In the following, we investigate the particle spectrum of heavy neutrinos, the extra $SU(2)_L$ doublet $\phi$ and the light neutrinos, coming from $\mathcal{L}_N$, $\mathcal{L}_{\phi^2}$ and $\mathcal{L}_y$, respectively.

\subsubsection{Heavy neutrinos $N$}

In Eq.~\eqref{eq:LN}, we do not describe the generation of heavy neutrino masses. Within the charge assignments, they can for example come from the
vacuum expectation values~(vevs) of additional scalar fields $\Phi_{1,2}$ of $U(1)_X$ charge $-2$ and $-4$, respectively. Furthermore, a Dirac mass term involving $N_1$ and $N_4$ can be written down because of their opposite $U(1)_X$ charges. Therefore, one has 
\begin{align}
	\mathcal{L}_N =  
	\frac{1}{2} \sum^3_{i=2}  \lee m_N \rii_{i}    N_i N_i +
	\frac{1}{2} \sum_{i,j=1,4}  \lee m_N \rii_{ij}   N_i N_j  + h.c.,
\end{align}
where all terms break the $U(1)_X$ symmetry except for the term $m_{41}N_4 N_1$, which conserves the $U(1)_X$ charge. The $4 \times 4$ heavy neutrino mass matrix therefore decomposes into two $2 \times 2$ blocks. Without lack of generality, the $N_2$ and $N_3$ mass matrix can be assumed diagonal, with mass eigenvalues $m_{N_2}$ and $m_{N_3}$. The $N_1$ and $N_4$ sector can be diagonalized as
\begin{align}
	\left(
		\begin{array}{c} N_4 \\ N_1 \\ \end{array}
	\right)_f =
	U_{41}
	\left(
		\begin{array}{c} N_4 \\ N_1 \\ \end{array}
	\right)_m, \quad 
	\text{with} \quad
	U_{41} = 
	\left(
  \begin{array}{cc}
    \cos\th_{41} & -\sin\th_{41} \, e^{i \al_{41}} \\
    \sin\th_{41} &  \cos\th_{41} \, e^{i \al_{41}} \\
  \end{array}
	\right),
\label{eq:N1-N4_mixing}
\end{align}
where the subscripts $f$ and $m$ refer to the flavour and mass basis, respectively. By convention we take $m_{N_1} < m_{N_4}$. We take the mixing angle $\theta_{41}$ and the phase $\alpha_{41}$ as free input parameters.

\subsubsection{Neutral components of $\phi$}
\label{sec:loop_computation}

Based on the quantum number assignment in Tab.~\ref{tab:quantum_number}, $\phi$ can interact up to quadratic terms as described in Eq.~\eqref{eq:Lphi},
\begin{align}
	\mathcal{L}_{\phi^2} \supset  m_\phi^{\prime 2} \phi^\dag \phi
  + c_1 (\tilde{H} \cdot H) (\phi^\dag \cdot\phi) 
	+ c_2 (\tilde{H} \cdot \phi) (H \cdot\phi^\dag),
\end{align}
where $\lan H \ran = (0, v/ \sqrt{2})^T$, $\lan \tilde{H} \ran = (v/ \sqrt{2}, 0 )^T$ and $\phi=\frac{1}{\sqrt{2}}(\phi^+_1 + i\phi^+_2, \phi^0_1 + i\phi^0_2 )^T$, in which $v$ is the Higgs vev and $\phi^{+,0}_{1,2}$ are real scalar fields.

\begin{figure}
\centering
\includegraphics[width=0.5\columnwidth]{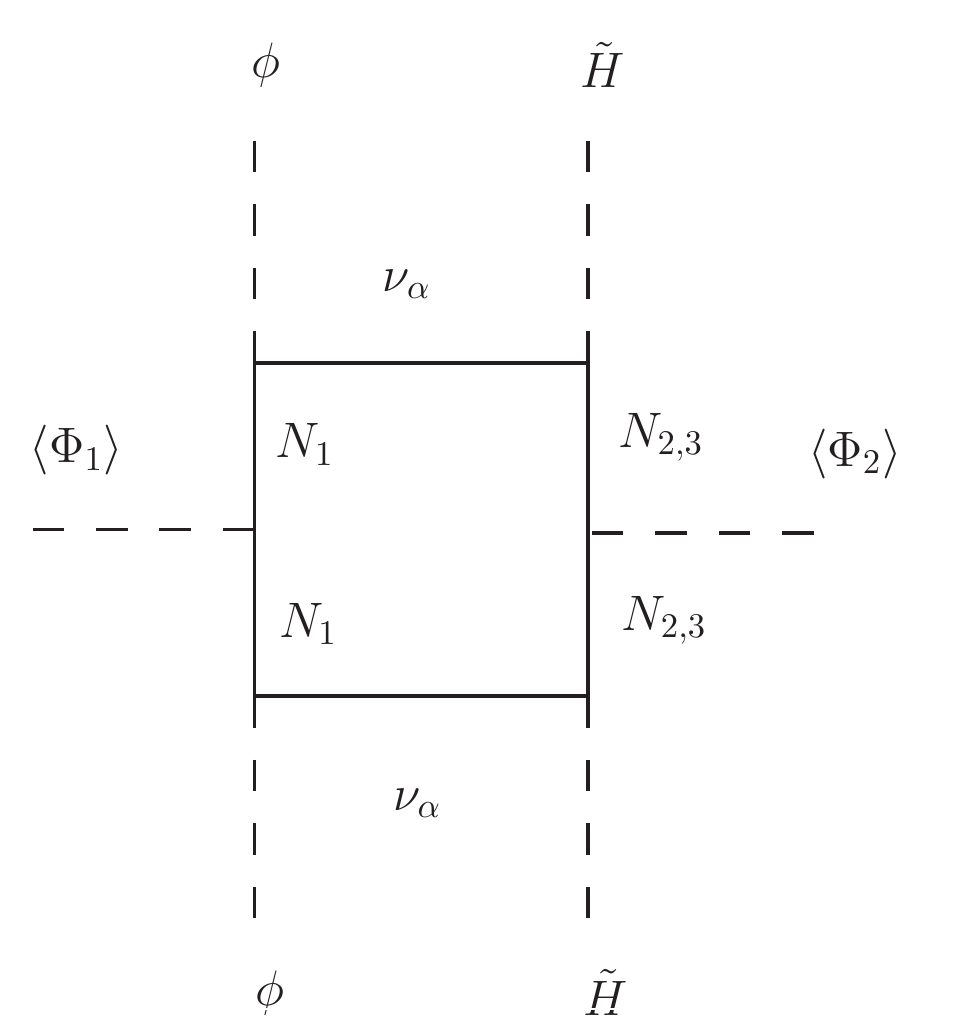}
\caption{Loop diagram contributing to the mass splitting $m^2_{\phi^0_1} - m^2_{\phi^0_2}$.}
\label{fig:phi_mass_splitting}
\end{figure}
As we shall see, successful generation of radiative light neutrino masses requires non-zero splitting $m^2_{\phi^0_1} - m^2_{\phi^0_2}$~\cite{Ma:2006km}, which will not arise from terms related to $c_1$ and $c_2$. On the other hand, the radiative corrections shown in Fig.~\ref{fig:phi_mass_splitting}, lift the degeneracy between $m^2_{\phi^0_1}$ and $m^2_{\phi^0_2} $ by introducing the term
\begin{align}
	\mathcal{L}_{\phi^2} \supset  
	c_3 (\tilde{H} \cdot \phi) (\tilde{H} \cdot \phi) + h.c.,
\label{eq:phi-deg}
\end{align}
which induces $\phi^0_1 -\phi^0_2$ mixing as well. To simplify the discussion, we start with zero $N_1-N_4$ mixing and we will generalize the result later with the help of Eq.~\eqref{eq:N1-N4_mixing}. Performing the computation of the loop in Fig.~\ref{fig:phi_mass_splitting}, the coefficient $c_3$ reads 
\begin{align}
	c_3(m_{N_1}) = - \sum_{j=2}^{3} \sum_{\alpha,\beta}
	\frac{ m_{N_1} m_{N_j} \lambda_{\alpha} \lambda_{\beta} y^*_{\alpha j} 
	y^*_{\beta j}}{8 \pi^2} \text{Re}(g_j),
\label{eq:c3_nm}
\end{align}
with the loop factor,
\begin{align}
	g_j &= C_0 (m_h^2, m_\phi^2, m_h^2, -m_\phi^2, m_{N_j}^2, 0, m_{N_1}^2) 
	\nonumber\\
	&- 2 m_{\phi}^2 D_0(m_h^2 - m_\phi^2, m_h^2, 4 m_\phi^2, m_\phi^2, m_\phi^2,
	m_h^2, m_{N_1}^2, m_{N_j}^2,0,0),
\end{align}
for which all external particles are put on-shell in the computation. The Higgs boson mass is indicated by $m_h$ and $C_0$, $D_0$ are the usual Passarino-Veltman integrals~\cite{Passarino:1978jh}\footnote{The imaginary part of the loop factor $g_j$ is $CP$-conserving since the antiparticles of $\phi$ and $H$ have the same loop factor. Thus it will not play a role in mass matrix diagonalization. This is the reason why we single out the real part only.}. The mass $m_\phi$ appearing in the integrals can simply be taken  to be either $m_{\phi^0_1}$ or $m_{\phi^0_2}$, as the mass difference is negligible in the loop function. The interaction eigenstates therefore mix and the resulting mass matrix is diagonalized by the transformation
\begin{align}
	\lee
		\begin{array}{c} \phi^0_1 \\ \phi^0_2  \\ \end{array}
	\rii_i
	=
	\lee
		\begin{array}{cc}
			 \cos\theta_\phi &  \sin\theta_\phi \\
			-\sin\theta_\phi &  \cos\theta_\phi \\
		\end{array}
	\rii
	\lee
		\begin{array}{c} \phi^0_1 \\ \phi^0_2 \\ \end{array}
	\rii_m,
\end{align}
where the subscripts $i$ and $m$ refer to the interaction and mass eigenstates, respectively. The mixing angle $\theta_\phi$ is purely determined by the contribution $c_3$, 
\begin{align}
	\tan 2\theta_\phi = \text{Im}(c_3) / \text{Re}(c_3)
\end{align}
and the mass eigenvalues are (we drop the subscript $m$ and implicitly refer to the mass eigenstates in the following),
\begin{align}
	m^2_{\phi^0_{1(2)}} &= m_\phi^{\prime 2} + \frac{c_1+c_2}{2} v^2 
	+(-) |c_3| v^2.
\end{align}
The resulting mass-squared difference is therefore simply given by
\begin{align}
	\Delta m^2_{\phi} \equiv m^2_{\phi^0_1} - m^2_{\phi^0_2} = 2 |c_3| v^2.
\label{eq:deltam_nm}
\end{align}
It is proportional to the small neutrino Yukawa couplings squared and consequently highly suppressed compared to the absolute mass values $m^2_{\phi_{1,2}}$.

Generalizing the result under the presence of the $N_1 - N_4$ mixing, $c_3$ consists of two contributions. The $N_1$ component is just Eq.~\ref{eq:c3_nm} multiplied by $\cos^2\theta_{41} e^{2 i \alpha_{41}}$ and the analogue second contribution due to $N_4$ is given by Eq.~\ref{eq:c3_nm} multiplied by $\sin^2\theta_{41}$ but with $m_{N_1}$ replaced by $m_{N_4}$,
\begin{align}
	c_3(m_{N_1},m_{N_4}) = \cos^2\theta_{41} e^{2 i \alpha_{41}} c_3(m_{N_1})  
	+ \sin^2\theta_{41} c_3(m_{N_4}).
\label{eq:deltam_wm}
\end{align}
Substituting Eq.~\ref{eq:deltam_wm} into Eq.~\ref{eq:deltam_nm}, one obtains the modified mass-squared difference.

\subsubsection{Light neutrinos}

%
\begin{figure}[t!]
\centering
\includegraphics[width=0.5\columnwidth]{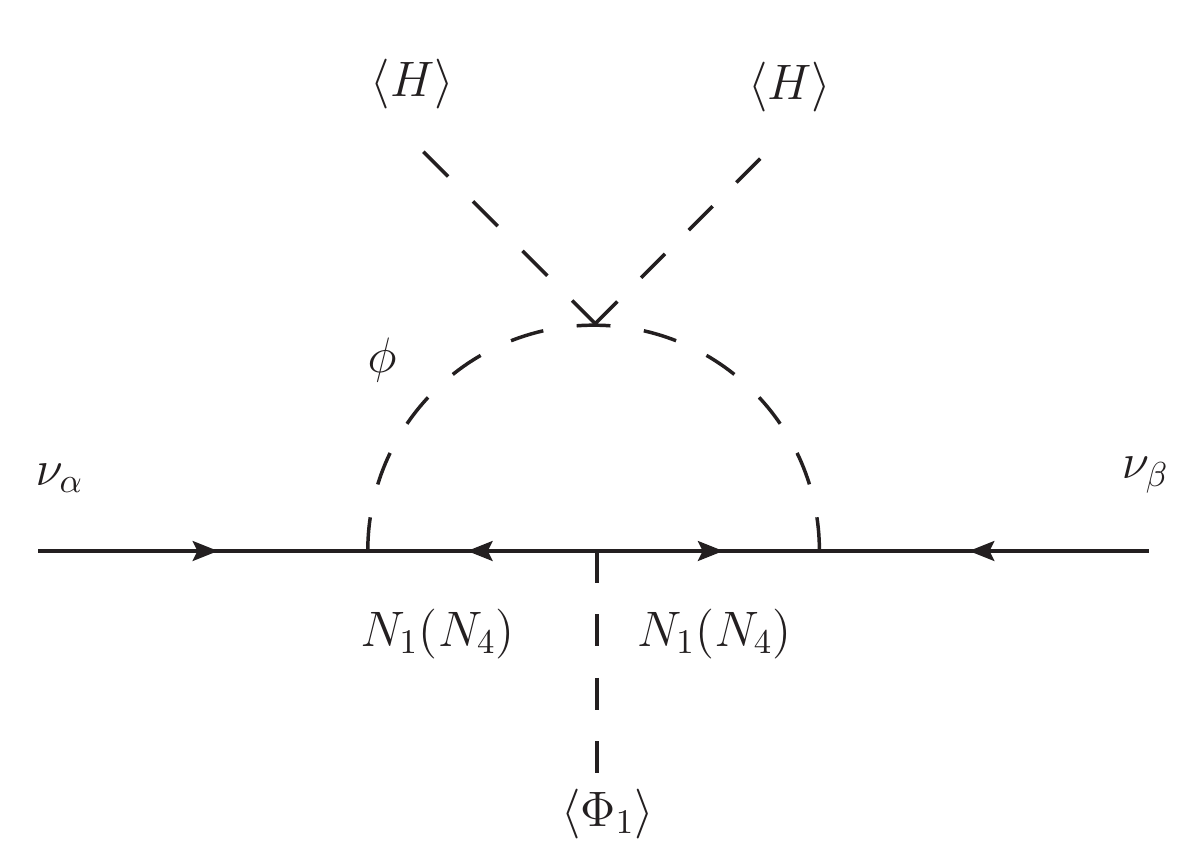}
\caption{Radiative contribution to light neutrino masses.}
\label{fig:loop}
\end{figure}
We here discuss the light neutrino mass matrix, by choosing the scenario $\eca$ for demonstration. The Lagrangian Eq.~\eqref{eq:Ly} in this case reads
\begin{equation}
	\mathcal{L} \supset 
	\sum_{\alpha=\mu,\tau} \sum_{i=2}^{3} y_{\alpha i } (L_\alpha \cdot H) N_i
	+ \sum_{\alpha=\mu,\tau} \la_\alpha (L_\alpha \cdot \phi) N_1
	+ \la_{N_4} (L_e \cdot \phi) N_4 + h.c. \, .
\end{equation}
The resulting full $7\times 7$ neutrino mass matrix in the flavour basis becomes,
\begin{align}
\label{eq:Mnu}
	\mathcal{M}_\nu=
  \begin{pmatrix}
    m_L   &  m_D  \\
    m_D^T & M   \\
  \end{pmatrix},
\end{align}
with the Dirac mass matrix $m_L$ arising from the original Yukawa interactions,
\begin{align}
	m_D = \frac{v}{\sqrt{2}} 
	\begin{pmatrix}
		0 & 0 & 0          & 0  \\
    0 & 0 & y_{\mu 2}  & y_{\mu 3} \\
    0 & 0 & y_{\tau 2} & y_{\tau 3} \\
  \end{pmatrix},
\label{eq:m_D}
\end{align}
and the heavy neutrino mass matrix 
\begin{align}
	M = \frac{v}{\sqrt{2}} 
	\begin{pmatrix}
		\lee m_{N} \rii_{11} & \lee m_{N} \rii_{41}   & 0  &0  \\
    \lee m_{N} \rii_{41} & \lee m_{N} \rii_{44} & 0 & 0 \\
      0 & 0 & m_{N_2} & 0 \\
      0 & 0 &  0 & m_{N_3} \\
  \end{pmatrix}.
\label{eq:M_44}
\end{align}
The left-handed mass matrix $m_L$ is not present at tree level but arises at loop level, cf. Fig.~\ref{fig:loop}. It can be expressed as
\begin{align}
	m_L = 
	\begin{pmatrix}
		\la_{N_4}^2         f_{44} & \la_{N_4} \la_\mu   f_{41} & \la_{N_4}   \la_{\tau}   f_{41} \\
    \la_{N_4} \la_\mu   f_{41} & \la_\mu^2           f_{11} & \la_\mu \la_{\tau}   f_{11} \\
		\la_{N_4} \la_{\tau} f_{41} & \la_\mu   \la_{\tau} f_{11} &         \la^2_{\tau} f_{11} \\
	\end{pmatrix},
\label{eq:m_L}
\end{align}
where the factors $f_{ij}$ refer to loop functions mediated by $N_1$ and $N_4$ in Fig.~\ref{fig:loop}. In the limit of the zero $N_1 - N_4$ mixing~\cite{Ma:2006km}, the loop functions are universally given by
\begin{align}
	f_0(m_{N_1}) &=
	\frac{m_{N_1}}{32 \pi^2} 
	\int_0^1 dx \ln \lee \frac{x m^2_{\phi^0_1} + (1-x) m^2_{N_1}}
	                          { x m^2_{\phi^0_2} + (1-x)m^2_{N_1} } \rii  \nonumber\\
	&= \frac{m_{N_1}}{32 \pi^2} 
	\lee \frac{m^2_{\phi^0_1}}{m^2_{\phi^0_1} - m^2_{N_1}} 
	\ln\lee \frac{ m^2_{\phi^0_1}}{m^2_{N_1}} \rii
	- \frac{m^2_{\phi^0_2}}{m^2_{\phi^0_2} - m^2_{N_1}} 
	\ln\lee \frac{ m^2_{\phi^0_2}}{m^2_{N_1}} \rii \rii.
\end{align}
To leading order in the small scalar $\phi$ mass splitting $\Delta m^2_\phi$ (cf. Eq.~\eqref{eq:deltam_nm}), it can be approximated as
\begin{align}
	f_0(m_{N_1}) \approx 
	\frac{m_{N_1}}{32\pi^2 (m^2_{\phi^0} - m^2_{N_1})^2}  
	\lee  m^2_{\phi^0} - m^2_{N_1}  
		- m^2_{N_1} \ln \lee \frac{m^2_{\phi^0}}{m^2_{N_1}} \rii 
	\rii
	\Delta m^2_\phi e^{- 2 i \theta_\phi}.
\label{eq:f_0_def}
\end{align}
Note that the radiative light neutrino mass generation requires both the lepton number violation from the Majorana masses $m_{N_{1,4}}$ and the $SU(2)_L$ symmetry breaking necessary to lift the degeneracy of the $\phi^0_1$ and $\phi^0_2$ masses, as described in Eq.~\eqref{eq:phi-deg}. Following the same procedure employed in Eq.~\eqref{eq:deltam_wm} to include the finite $N_1 - N_4$ mixing, the individual loop functions appearing in Eq.~\eqref{eq:m_L} are expressed as
\begin{align}
\label{eq:f11_f41_f44}
	f_{11} &= \cos^2\theta_{41} e^{2i\alpha_{41}} \, f_0(m_{N_1}) 
		+ \sin^2\theta_{41} \, f_0(m_{N_4}), \nonumber\\
	f_{44} &= \sin^2\theta_{41} e^{2i\alpha_{41}} \, f_0(m_{N_1}) 
		+ \cos^2\theta_{41} \, f_0(m_{N_4}), \\
	f_{41} &= \frac{1}{2}\sin(2\theta_{41}) \left(-e^{2i\alpha_{41}} \, f_0(m_{N_1}) 
		+  f_0(m_{N_4}) \right). \nonumber
\end{align}
The light neutrino mass matrix can now be calculated from Eq.~\eqref{eq:Mnu}, $m_\nu = m_L - m_D M^{-1} m_D^T$. It is diagonalized by the charged-current mixing matrix $U$,
\begin{align}
	\text{diag}(m_{\nu_1}, m_{\nu_2}, m_{\nu_3}) = U^T (m_L -m_D M^{-1} m_D^T) U.
\label{eq:diago}
\end{align}
The charged current diagonalization matrix $U$ is given in the standard parametrization by
\begin{align}
	U =
	&
	\begin{pmatrix}
		1 &      0  &       0 \\
		0 & c_{23}  & -s_{23} \\
		0 & s_{23}  &  c_{23}
	\end{pmatrix}
	\cdot
	\begin{pmatrix}
		c_{13}                  & 0 & -s_{13} e^{-i\delta} \\
		0                       & 1 &           0          \\
		s_{13} e^{i\delta} & 0 &  c_{13}
	\end{pmatrix} \cdot
	\begin{pmatrix}
		c_{12} & -s_{12} & 0 \\
		s_{12} &  c_{12} & 0 \\
		     0 &       0 & 1
	\end{pmatrix} \cdot
	\begin{pmatrix}
		1 & 0              & 0              \\
		0 & e^{i \alpha_1} & 0              \\
		0 & 0              & e^{i \alpha_2}
	\end{pmatrix},
	\label{eq:Umax}
\end{align}
with the usual mixing angles $\theta_{12}$, $\theta_{13}$, $\theta_{23}$ ($c_{ij} \equiv \cos\theta_{ij}$, $s_{ij} \equiv \sin\theta_{ij}$), the Dirac $CP$ phase $\delta$, and the Majorana phases $\alpha_1$, $\alpha_2$.

\section{Fitting the Neutrino Data}
\label{fitting_nu_data}

We are now in a position to calculate the phenomenological consequences of the hybrid neutrino mass model described above. We will first use the neutrino oscillation data to fix some of the model parameters. Subsequently, we will analyze the viability of the model with respect to LFV processes and its potential to predict the observed relic abundance with the lightest heavy neutrino $N_1$ as the DM candidate.

We stay within the $\eca$ scenario with $X_{L_e}=0$ and $X_{L_\mu}=X_{L_\tau}=2$ to demonstrate the procedure. From Eq.~\eqref{eq:m_L} and \eqref{eq:m_D}, the $3\times 3$ light neutrino mass matrix is given by
\begin{align}
	m_\nu=
	\begin{pmatrix}
    \la_{N_4}^2 f_{44}  & \la_{N_4} \la_\mu f_{41}  & \la_{N_4} \la_\tau f_{41}   \\
    \la_{N_4} \la_\mu f_{41}  & \la^2_\mu f_{11}  - \frac{y^2_{\mu 2}}{2m_{N_2}} v^2 -  \frac{y^2_{\mu 3}}{2m_{N_3}} v^2 & 
   \la_\mu \la_\tau f_{11} - \frac{y_{\mu 2} y_{\tau 2}}{2m_{N_2}} v^2 -  \frac{y_{\mu 3} y_{\tau 3}}{2m_{N_3}} v^2   \\
		\la_{N_4} \la_\tau f_{41} &  \la_\mu \la_\tau f_{11} - \frac{y_{\mu 2} y_{\tau 2}}{2m_{N_2}} v^2 -  \frac{y_{\mu 3} y_{\tau 3}}{2m_{N_3}} v^2 &
		\la^2_\tau f_{11}  - \frac{y^2_{\tau 2}}{2m_{N_2}} v^2 -  \frac{y^2_{\tau 3}}{2m_{N_3}} v^2 \\
	\end{pmatrix}.
	\label{eq:eca_mnu}
\end{align}
 \begin{table}[t!]
\centering
\begin{tabular}{c c c c c c c c}
\hline
& $\sin^2 2\theta_{12}$ &  $\sin^2 2\theta_{23}$ & $\sin^2 2\theta_{13}$ & 
  $\Delta m^2_{sol}$ (eV$^2$) & $|\Delta m^2_{atm}|$ (eV$^2$)  \\
\hline
best-fit & 0.857 & 1.00 & 0.095 & $ 7.50\times 10^{-5}$ & $ 2.32\times 10^{-3}$  \\
$1\sig$  & 0.024 & 0.30 & 0.010 & $ 2.00\times 10^{-6}$ & $ 1.00\times 10^{-4}$  \\
\hline
\end{tabular}
\caption{Best-fit values and $1\sigma$ standard deviations of neutrino oscillation parameters used in this work. The values are taken from Refs.~\cite{Beringer:1900zz, Abe:2013fuq, Ade:2013zuv}. We neglect small asymmetries in uncertainties and differences between the HN and IH cases as present in the recent global fits~\cite{Forero:2014bxa,Gonzalez-Garcia:2014bfa}.}
\label{tab:observables}
\end{table}
As we will discuss in detail below, the couplings $\lambda_\mu$, $\lambda_\tau$ and $\la_{N_4}$ are not only responsible for the radiative neutrino masses but also generate charged LFV process, as shown in Fig.~\ref{fig:mu_to_ega}. Large $\lambda_\mu$, $\lambda_{N_4}$ imply large rates for the decay $\mu\to e \gamma$, which is constrained by $\text{Br}(\mu \rightarrow e \ga)< 5.7 \times 10^{-13}$~\cite{Sawada:2014uba}.
In order to reproduce the neutrino oscillation observables shown in Tab.~\ref{tab:observables}, we should equate Eq.~\eqref{eq:eca_mnu} to the neutrino mass matrix,
$m^{\text{obs}}_{\nu}$, determined by the neutrino masses and the neutrino mixing matrix\footnote{In addition to the observables, one still needs to fix $m_{\nu_1}~(m_{\nu_3})$, $\delta$ and $\al_{1,2}$ in the normal hierarchy~(NH)~(inverted hierarchy~(IH)) case.}. Note that in the $\eca$ case, one has the following free parameters: the masses of the heavy particles, $m_\phi$ and $m_{N_i}$, four Yukawa couplings $y_{\al i}$, the couplings $\lambda_{N_4}$, $\lambda_\mu$, $\lambda_\tau$, and the $N_1-N_4$ mixing angle $\th_{41}$ and phase $\al_{41}$. Given the experimental information on the light neutrino mass matrix, several parameters can be fixed. We use the following procedure to determine the values of $y_{\al i}$, $\la_{\mu}$, $\th_{41}$ and $\al_{41}$:     
\begin{itemize}
\item The ratio of $ \lee m^{\text{obs}}_{\nu} \rii_{12}$ to $ \lee m^{\text{obs}}_{\nu} \rii_{13}$ is $\la_{\mu} \la_{N_4} f_{41}/ (\la_{\tau} \la_{N_4} f_{41})$. Therefore, $\la_{\mu}$ is simply $\la_{\tau} \lee m^{\text{obs}}_{\nu} \rii_{12}/\lee m^{\text{obs}}_{\nu} \rii_{13}$.  
\item The ratio of $ \lee m^{\text{obs}}_{\nu} \rii_{11}$ to $ \lee m^{\text{obs}}_{\nu} \rii_{13}$ is $\la_{N_4}f_{44}/(\la_{\tau}f_{41})$. From Eq.~\eqref{eq:f_0_def} and \eqref{eq:f11_f41_f44}, $f_{44}/f_{41}$ depends only on $\th_{41}$ and $\al_{41}$ but not $y_{\al i}$. Hence, one can solve for
$\th_{41}$ and $\al_{41}$ from $ \lee m^{\text{obs}}_{\nu} \rii_{11}/\lee m^{\text{obs}}_{\nu} \rii_{13}$.
\item Once $\la_{\mu}$, $\th_{41}$ and $\al_{41}$ are known, one can employ $ \lee m^{\text{obs}}_{\nu} \rii_{13}$, $ \lee m^{\text{obs}}_{\nu} \rii_{22}$,
$ \lee m^{\text{obs}}_{\nu} \rii_{23}$ and $ \lee m^{\text{obs}}_{\nu} \rii_{33}$ to solve for four $y_{\al i}$ in the context of quadratic equations. In general, one has 16 different solutions of $y_{\al i}$.
\end{itemize}

\begin{figure}[t]
\centering
\includegraphics[width=0.48\columnwidth]{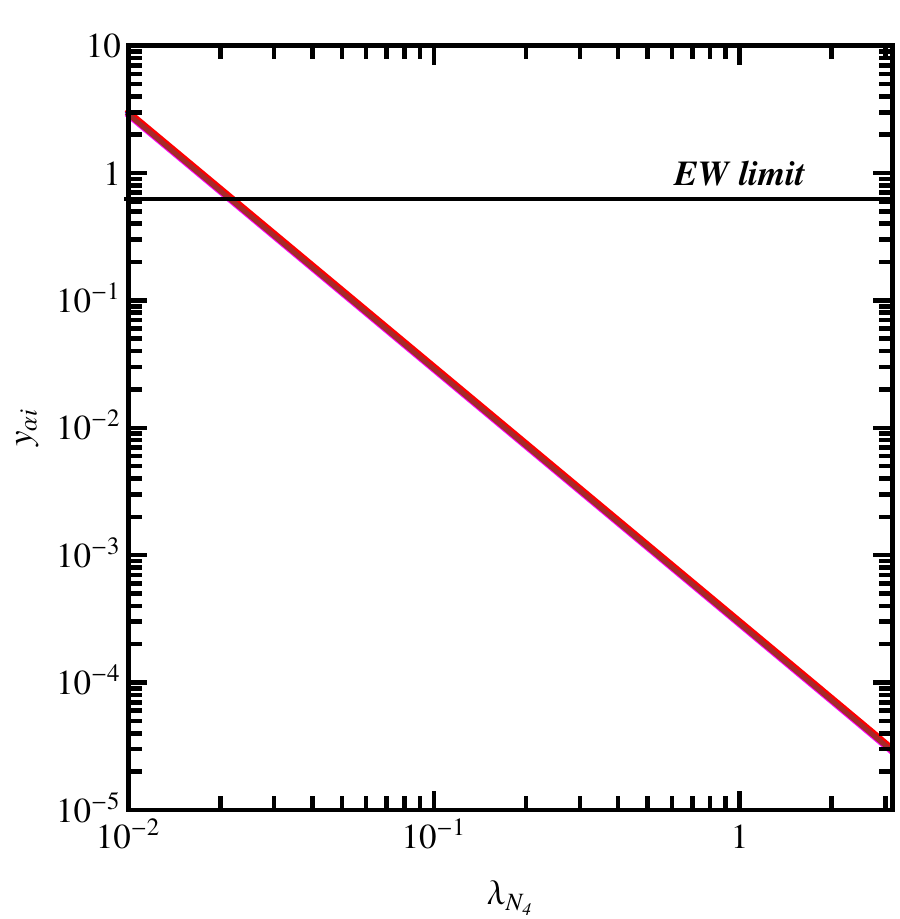}
\includegraphics[width=0.48\columnwidth]{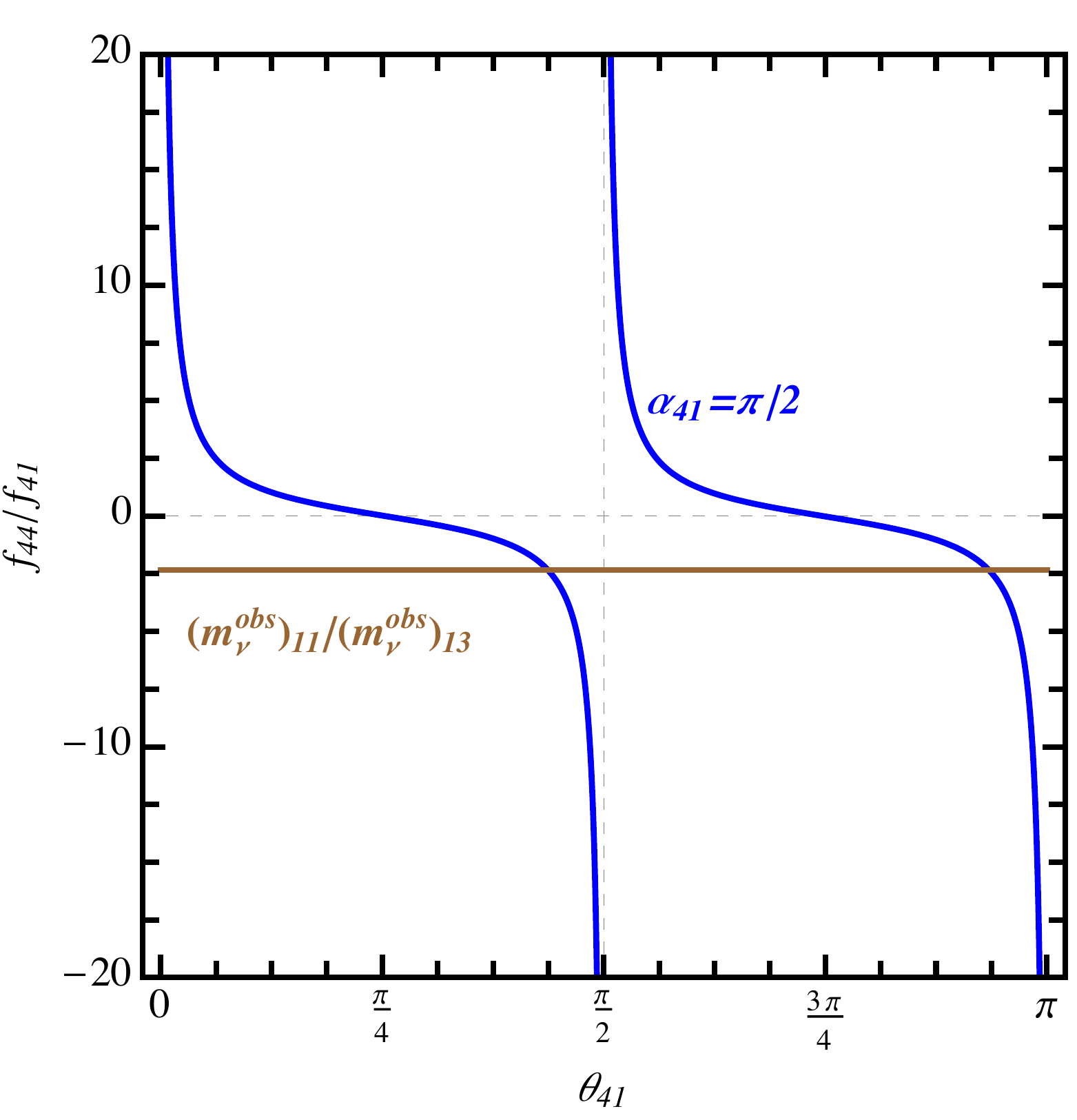}
\caption{Left: Yukawa couplings $y_{\al i}$ as a function of $\la_{N_4} = \la_\tau$ with the remaining parameters set as in the benchmark scenario of Tab.~\ref{tab:m_solve_yla}. The different Yukawa couplings are essentially degenerate. Right: Ratio of $f_{44}$ to $f_{41}$ as a function of $\th_{41}$ for $\al_{41} = \pi/2$~(blue), where the benchmark point in Tab.~\ref{tab:m_solve_yla} is assumed. Note that $|f_{44}/f_{41}|$ is of $\mathcal{O}(10^5)$ for $\al_{41}=0$ and is beyond the scope of the plot since $m_{N_1}\approx m_{N_4}$ leading to a vanishing value of $f_{41}$.}
\label{fig:y_la_sol}
\end{figure}
\begin{table}[b]
\centering
\begin{tabular}{ccccccccccccc}
\hline
$m_\text{lightest}$ & $\delta$ & $\alpha_1$ & $\alpha_2$ & 
$\lambda_{N_4}$ & $\lambda_\tau$ &
$m_\phi$ & $m_{N_1}$ & $m_{N_4}$ & $m_{N_2}$ & $m_{N_3}$ \\
$ 10^{-2}$~eV & 0 & 0 & 0 & 0.5 & 0.5 &
1.2 TeV & 1 TeV & 1.01 TeV & 2 TeV & 2.1 TeV  \\
\hline
\end{tabular}
\caption{Reference scenario used in our numerical analysis.}
\label{tab:m_solve_yla}
\end{table}
In Fig.~\ref{fig:y_la_sol}, with $\la_{\tau}=\la_{N_4}$, we present values of $y_{\al i}$ as a function of $\la_{N_4}$ with the benchmark parameters listed in Tab.~\ref{tab:m_solve_yla}. It is clear that $y_{\al 2} \approx y_{\al 3}$ for the degenerate spectrum, $m_{N_2}\sim m_{N_3}$. The upper bound on $|y_{\al i}| \lesssim 0.6$ results from the LFV constraints via $N_{2,3}$ radiative contributions and the EW precision data~\cite{Atre:2009rg}. The radiative corrections to neutrino masses induced by $N_1$ and $N_4$ are roughly $\la^2/(16\pi^2)^2$ times the seesaw type-I contributions from Eq.~\eqref{eq:c3_nm}, where the factor $1/(16\pi^2)^2$ is due to the second loop order involved. Generally speaking, due to sizable $\th_{13}$ mixing, the loop contributions have to be of the same order as the seesaw type-I tree-level contributions. This implies that the $\la$ couplings have to be large or $y_{\al i}$ have to be large with fine-tuning within the seesaw type-I contributions. It turns out that large $\la$ values~($\gtrsim 1 $) lead to large DM annihilation cross sections, thereby suppressing the DM relic density. Therefore, we have to resort to large $y_{\al i}$, which can be easily realized if the inverse-seesaw~\cite{Mohapatra:1986bd} is involved in the context of $m_{N_2} \simeq m_{N_3}$. As shown in Fig.~\ref{fig:y_la_sol}~(left), all $y_{\alpha i}$ are of the same order and are much larger than the naive seesaw estimate $y \approx 10^{-6}$ for TeV scale $N$. As mentioned above, $\th_{41}$ and $\al_{41}$ can be solved from the ratio of $ \lee m^{\text{obs}}_{\nu} \rii_{11}$ to $\lee m^{\text{obs}}_{\nu} \rii_{13}$~($\sim \la_{N_4}f_{44}/(\la_{\tau}f_{41})$). The solution, however, is not unique. In the limit of $\la_{N_4}=\la_{\tau}$, the ratio becomes
\begin{align}
\frac{ \lee m^{\text{obs}}_{\nu} \rii_{11}}{ \lee m^{\text{obs}}_{\nu} \rii_{13}}
=\frac{ \sin^2\theta_{41} e^{2i\alpha_{41}}  
		+ \cos^2\theta_{41} \, f_0(m_{N_4})/f_0(m_{N_1}) }{ -\cos\theta_{41}\sin\theta_{41} e^{2i\alpha_{41}} 
		+ \cos\theta_{41}\sin\theta_{41} \, f_0(m_{N_4})/f_0(m_{N_1}) },
\end{align}
where again $ f_0(m_{N_4})/f_0(m_{N_1})$ depends on $m_{N_1}$ and $m_{N_4}$ but not on $y_{\al i}$. As shown in Fig.~\ref{fig:y_la_sol}~(right) with the reference scenario in Tab.~\ref{tab:m_solve_yla}, for $\al=\pi/2$, there are two distinctive solutions located in the region of $(0,\pi/2)$ and $(\pi/2, \pi)$, respectively. By contrast, for  $\al=0$, there exists no solution due to $m_{N_1}\approx m_{N_4}$ and in turn vanishing $f_{41}$~(and very large $|f_{44}/f_{41}|$). Therefore, the physical range for $\th_{41}$ ranges from 0 to $\pi$, which is the same as $\al_{41}$. In this work, we restrain ourselves on the region of $\th_{41} \in  (0,\pi/2)$.

\section{Lepton Flavour Violation}
\label{sec:LFV}

%
\begin{figure}[t!]
\centering
\includegraphics[width=0.5\columnwidth]{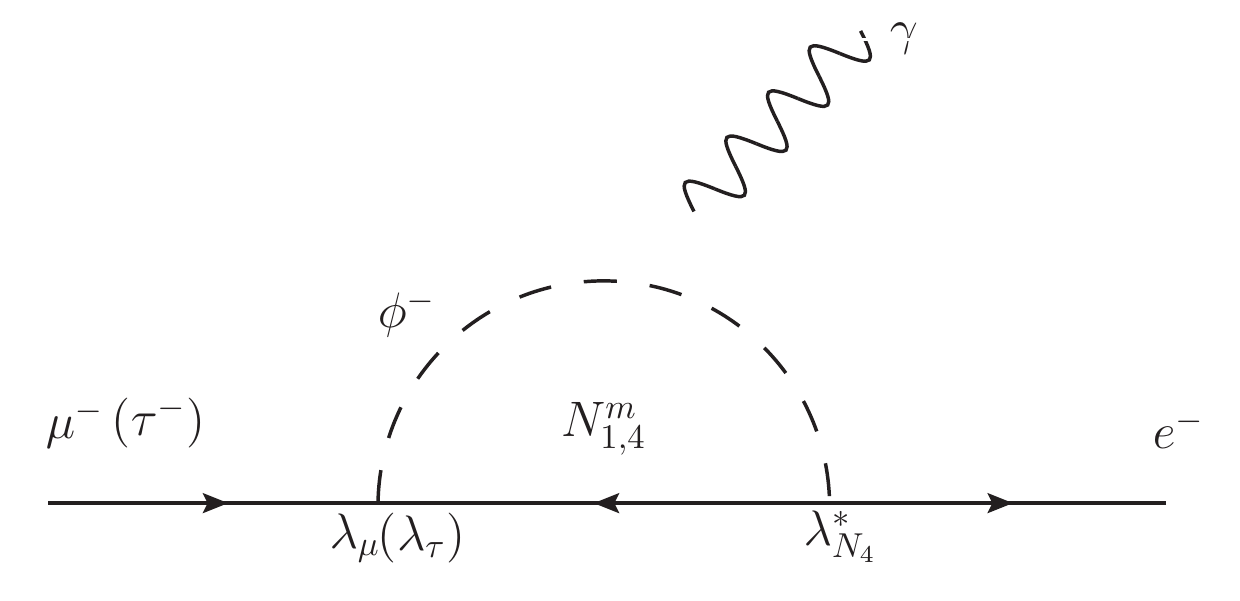}
\caption{Loop diagram mediating the LFV processes $\ell_\alpha\to \ell_\beta\gamma$.}
\label{fig:mu_to_ega}
\end{figure}
In this Section, we discuss the LFV processes $ \ell_\alpha \to \ell_\beta \gamma$ induced by the $N_1-N_4$ mixing. We stick to the $\eca$ case but the generalization to the other cases is straightforward and will be discussed below. The three LFV processes $\tau \to \mu \gamma$, $\tau \to e \gamma$ and $\mu \to e \gamma$ are proportional to $|\la_{\tau}\la_{\mu}|^2$, $|\la_{\tau}\la_{N_4}|^2$ and $|\la_{\mu}\la_{N_4}|^2$, respectively, when mediated by $N_1$ and $N_4$, cf. Fig.~\ref{fig:mu_to_ega}. Note that $\tau \to e \gamma$ and $\mu \to e \gamma$ vanish if the $N_1-N_4$ mixing angle $\theta_{41}$ is zero, whereas $\tau \to \mu \gamma$ is non-zero even for $\theta_{41}=0$ since it can be induced by purely $N_1$. 

The standard seesaw type-I contributions to LFV processes mediated by $N_2$ and $N_3$ together with a SM $W$ in the loop are of course also present. In the $\tca$ case, only the $\mu$ and $\tau$ lepton couple to $N_{2,3}$, so only $\tau \to \mu \ga$ will be induced. The corresponding branching fraction~\cite{Deppisch:2010fr} is, in the limit of $m_{N_2}\approx m_{N_3}~(\equiv m_N)$,
\begin{align}
\label{eq:BllgammaApprox}
	\text{Br}(\tau\to \mu\gamma)\ &\approx \
	1.5\cdot 10^{-4}\times 
      \frac{v^4}{ m_N^4}\  G^2_\gamma(m^2_N/m^2_W) \; \left|  \sum^3_{i=2} \lee y^*\rii_{\tau i} \lee y\rii_{ \mu i} \right|^2,
\end{align}
where
\begin{align}
	\label{eq:Ggamma}
	G_\gamma(x)\ =\ 
	-\ \frac{2x^3+5x^2-x}{4(1-x)^3}-\frac{3x^3}{2(1-x)^4}\ln x \;.
\end{align}
The current limit $\text{Br}(\tau \to \mu\gamma) \approx 10^{-8}$ corresponds to $\sqrt{y_{\tau i}y_{\mu i}} \sim 0.5$. As we shall see later, the model is also constrained from the $\tau \to \mu \ga$ limit in addition to  $\tau~(\mu) \to e \ga$ induced by $N_{1,4}$. 

In our computation of the LFV rates, we set the final state lepton mass to zero. The $\ell_\alpha \to \ell_\beta \gamma$ decay widths are given by (setting the final state lepton mass to zero) 
\begin{align}
	\Gamma_{\tau(\mu) \to e \gamma} &= 
	\frac{e^2 \cos^2\theta_{41} \sin^2\theta_{41} m^5_{\tau(\mu)}}{16^3 \pi^5} 
	\left\vert
		\la_{\tau(\mu)} \la^*_{N4}  (f(m_{N_1}) - f(m_{N_4})) 
	\right\vert^2,  \\
	\Gamma_{\tau \to \mu \gamma} &=
	\frac{e^2 m^5_\tau}{16^3 \pi^5}
	\left\vert
		\la_{\tau} \la^*_{\mu} (\cfot f(m_{N_1}) + \sfot f(m_{N_4})) 
	\right\vert^2,
\label{eq:tau_e_ga}
\end{align}
with
\begin{align}
	f(m_{N_i}) = \frac{1}{6(m^2_{N_i} - m^2_{\phi})^4} 
	\lee
			2 m^6_{N_i} + 3 m^4_{N_i} m^2_{\phi} - 6 m^2_{N_i} m^4_{\phi} + m^6_{\phi}
		- 6 m^4_{N_i} m^2_{\phi} \log \lee \frac{m^2_{N_i}}{m^2_\phi} \rii
	\rii.
\end{align}
Note that there is a strong suppression of $\tau(\mu) \to e\gamma$ for $m_{N_4} \approx m_{N_1}$, while $\tau \to \mu \ga$ is unsuppressed. In the limit of $\Delta m_{41} \equiv m_{N_4} - m_{N_1} \ll m_{N_1} \approx m_{\phi}$, the $\tau$- and $\mu$-related LFV decay widths become
\begin{align}
	\Gamma_{\tau (\mu) \to  e \gamma} &\approx 
	\frac{e^2 |\la_{\tau(\mu)} \la_{N_4}|^2 \sin^2 2\theta_{41}}{16^3 \pi^5}
	\frac{m^5_{\tau(\mu)} (\Delta m_{41})^2}{900 m^6_{\phi}}, \\
	\Gamma_{\tau \to \mu \gamma} &\approx 
	\frac{e^2 |\la_{\tau} \la_{\mu}|^2}{16^3 \pi^5}
	\frac{m^5_{\tau}}{144 m^4_{\phi}}.
\end{align}
The corresponding decay branching fractions are
\begin{align}
	\text{Br}(\tau \to e \gamma) &\approx
	6.4 \times 10^{-16} \sin^2 2\theta_{41} |\la_{\tau} \la_{N_4}|^2  
	\lee \frac{\Delta m_{41}}{\text{GeV}} \rii^2 
	\lee \frac{\text{TeV}}{m_\phi}        \rii^6, \nn\\
	\text{Br}(\mu \to e \gamma) &\approx
	3.6 \times 10^{-15} \sin^2 2\theta_{41} |\la_{\mu} \la_{N_4}|^2  
	\lee \frac{\Delta m_{41}}{\text{GeV}} \rii^2 
	\lee \frac{\text{TeV}}{m_\phi}        \rii^6, \nn\\
	\text{Br}(\tau \to \mu \gamma) &\approx
	4.0 \times 10^{-9}  |\la_{\mu} \la_{\tau}|^2  
	\lee \frac{\text{TeV}}{m_\phi}        \rii^4.
	\label{eq:br_sim}
\end{align}
They can be compared to the experimental limits $\text{Br}(\tau~(\mu) \to e \gamma) < 3.3~(4.4) \times 10^{-8}$~\cite{Beringer:1900zz}  and $\text{Br}(\mu \to e \gamma)<  5.7 \times \times 10^{-13}$~\cite{Sawada:2014uba}. Roughly speaking, for TeV scale $m_\phi$ and $\la \sim 1$, the loop-induced transition $\text{Br}(\tau \to \mu \ga)$ is slightly below the current bound, while $\text{Br}(\mu \to e \ga)$ violates the current limit for $\Delta m_{41} \gtrsim 10$ GeV.

\section{DM Relic Abundance}
\label{sec:DM}

%
\begin{figure}
\centering
\includegraphics[width=0.8\columnwidth]{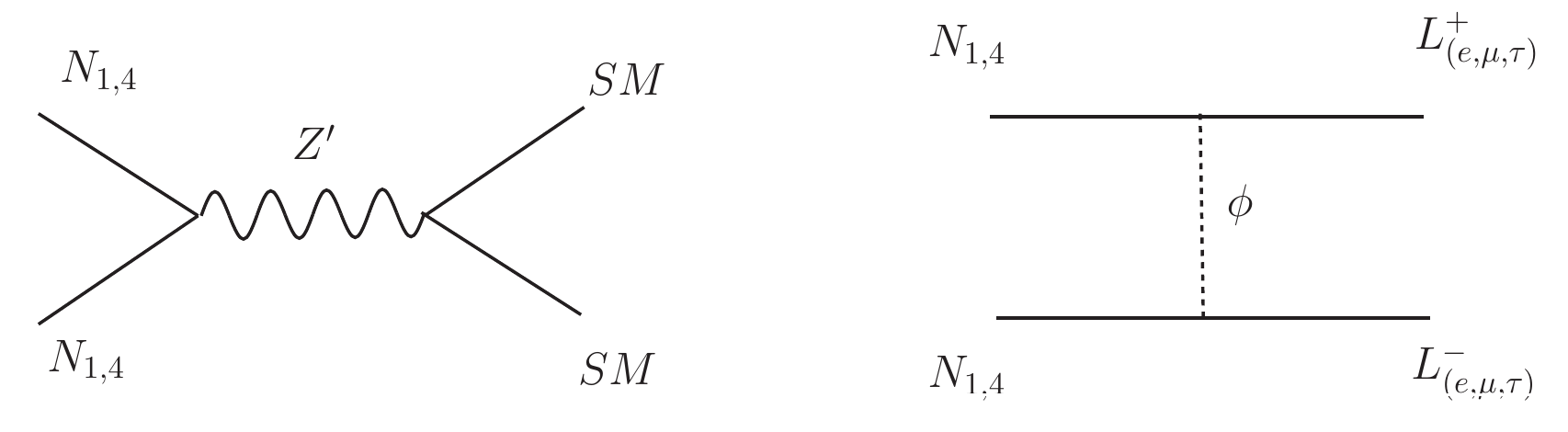}
\caption{DM annihilation via $s$-channel $Z'$ and $t$-channel $\phi$ exchange.}
\label{fig: DM_density}
\end{figure}
In this Section, we compute the DM relic abundance, including co-annihilations. The main processes are mediated by $Z^\prime$ and $\phi$ exchange as shown in Fig.~\ref{fig: DM_density}. In this work, we concentrate on the interplay between the DM particle $N_1$ and the light neutrinos and we study the effect of $\phi$ exchange in detail. Therefore, the observed DM relic density $\Omega_{\text{DM}} h^2 = 0.120 \pm 0.003$ is used as a lower bound since the inclusion of $Z^\prime$ exchange will only further decrease the DM density. In other words, for scenarios with $\Omega_{\text{DM}} h^2 > 0.12$, the DM abundance can be reduced to the observed value with the help of $Z^\prime$ exchange processes whereas points with $\Omega h^2<0.12$ will be excluded regardless of $Z^\prime$ exchange contributions.

\begin{figure}
\centering
\includegraphics[width=0.7\columnwidth]{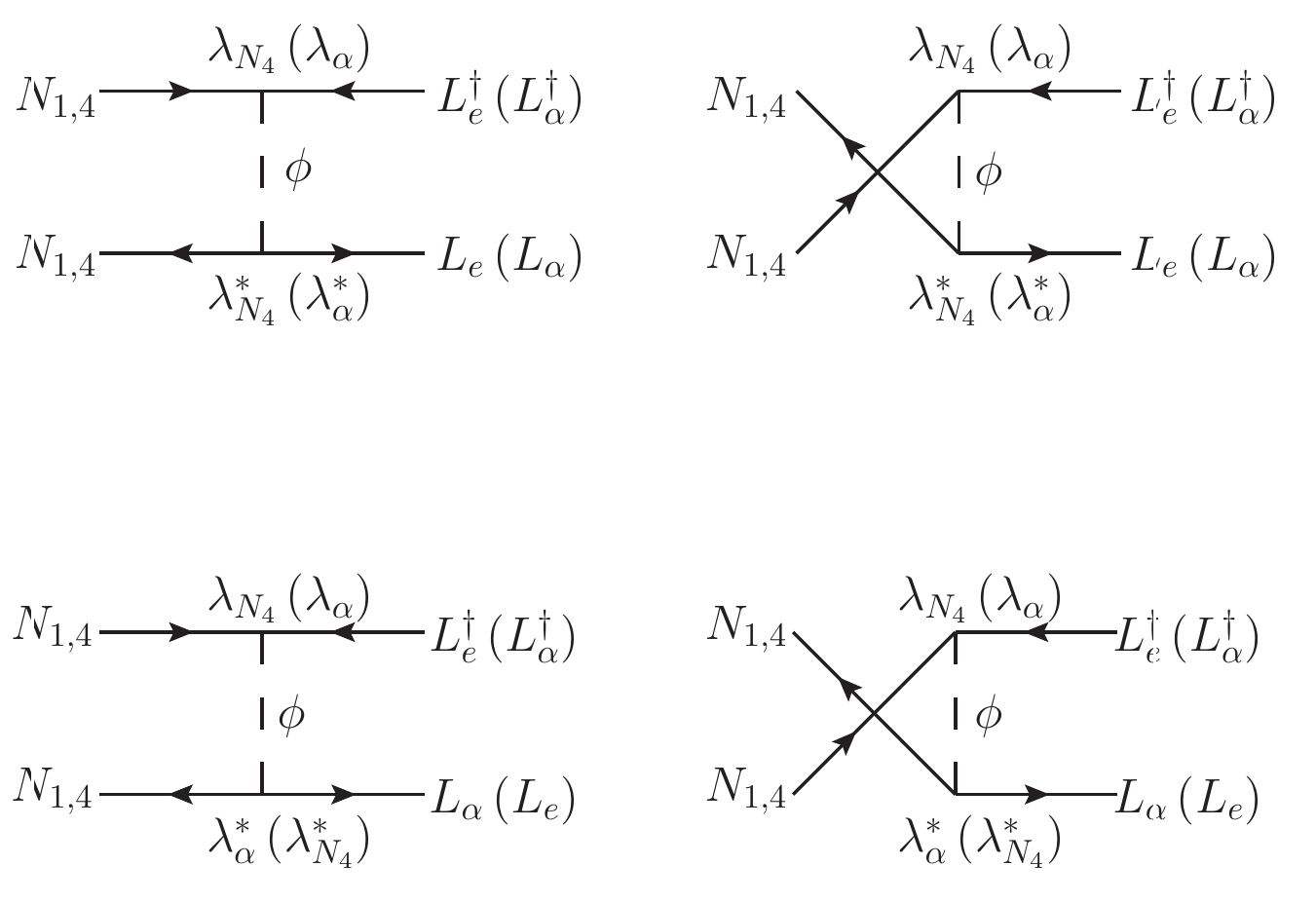}
\caption{All $\phi$ exchange (co-)annihilation channels in the $\eca$ scenario with $\al=\mu,\tau$.}
\label{fig: NN_LL}
\end{figure}

We again choose the $\eca$ case for demonstration. The relevant (co-)annihilation processes are shown in Fig.~\ref{fig: NN_LL}, where $N_{1,4}$ refer to the mass eigenstate and $L$ represents the $SU(2)_L$ doublet $(\nu_L , e^-_L)^T$. The upper panels feature the same lepton flavour final states, while the lower panels give rise to lepton flavour violation. Note that there exists a relative minus sign between $t$- and $u$- channel due to the odd permutation on the initial state fermions. We collect all (co-)annihilation cross sections in Appendix~\ref{sec:sigmav}. Because we are mainly interested in the degenerate region of $m_{N_1}\sim m_{N_4}$, in order to simplify the computation, we assume $m_N=\lee m_{N_1} + m_{N_4}\rii/2$ for co-annihilation processes. Our computation of the relic density under the presence of co-annihilations can be found in Ref.~\cite{Griest:1990kh}. Note that loop-induced LFV processes and $\phi$ exchange (co)-annihilations are correlated through the same interacting vertices, establishing the interplay between LFV and DM phenomenology. Finally, we would like to point out that our DM candidate, $N_1$, is a Majorana particle and the mass difference between $N_1$ and $N_4$ are assumed to be much larger than keV, which is the typical scale of the momentum transferred in DM direct detection experiments. The DM-nucleon cross section is either spin-dependent or velocity suppressed, so we will not consider the DM direct search bounds here. 

We conclude this section with a comment on the DM annihilation via $Z^\prime$ exchange. The corresponding annihilation cross section is of the same order as the $\phi$ exchange ones if $\la \sim g_{Z^\prime}$ and $m_{Z^\prime} \sim m_{\phi}$, where $g_{Z^\prime}$ is the $U(1)_X$ gauge coupling constant. Nonetheless, the $Z^\prime$ exchange processes are $s$-channel and can have the resonant enhancement if $2m_{N_1} \sim m_{Z^\prime}$. In this case, it will be larger than the $\phi$ exchange cross section roughly by a factor of $\lee 16 \pi/g^2_{Z^\prime} \rii^2$ and become the dominant contribution to the DM annihilation. 
%

\section{Results}
\label{sec:result}

As discussed in Section~\ref{fitting_nu_data}, we start analyzing the phenomenology of our model by reducing the number of free parameters through fitting the light neutrino oscillation data. Unless otherwise noted, we use the best-fit values for the oscillation parameters as listed in Tab.~\ref{tab:observables}, where we neglect differences between the NH and IH cases for simplicity. As a result, we can express the Yukawa couplings $y_{\alpha i}$, ($\alpha=e,\mu$, $i=1,2$), the coupling $\lambda_\mu$ and the mixing parameters $\theta_{41}$, $\alpha_{41}$ in terms of the couplings $\lambda_{N4}, \lambda_\tau$ and the masses of the heavy states ($m_{N_i}$, $m_\phi$), for a given light neutrino mass matrix. The latter is described by the lightest neutrino mass $m_\text{lightest}$, the hierarchy (NH, IH) and the $CP$ phases $\delta$, $\alpha_1$, $\alpha_2$, in addition to the best-fit oscillation parameters. We restrict ourselves to the $CP$ conserving case, and choose the $CP$ phases to be either $\delta = 0, \alpha_1 = \alpha_2 = 0$ or $\delta = 0, \alpha_1 = \alpha_2 = \pi/2$. We also do not attempt to provide a comprehensive scan over the whole parameter space including all the degeneracies discussed in Section~\ref{fitting_nu_data}. Instead we use the simplification $\lambda_{N_4} = \lambda_\tau$, especially to discuss the physically interesting parameter region where the $\lambda$ couplings are of order one. As we have seen in the previous sections, this corresponds to potentially large LFV, a viable DM relic density and Yukawa couplings $y_{\alpha i} \approx 10^{-3} - 10^{-2}$.

\begin{figure}[t]
\centering
\includegraphics[width=0.45\textwidth]{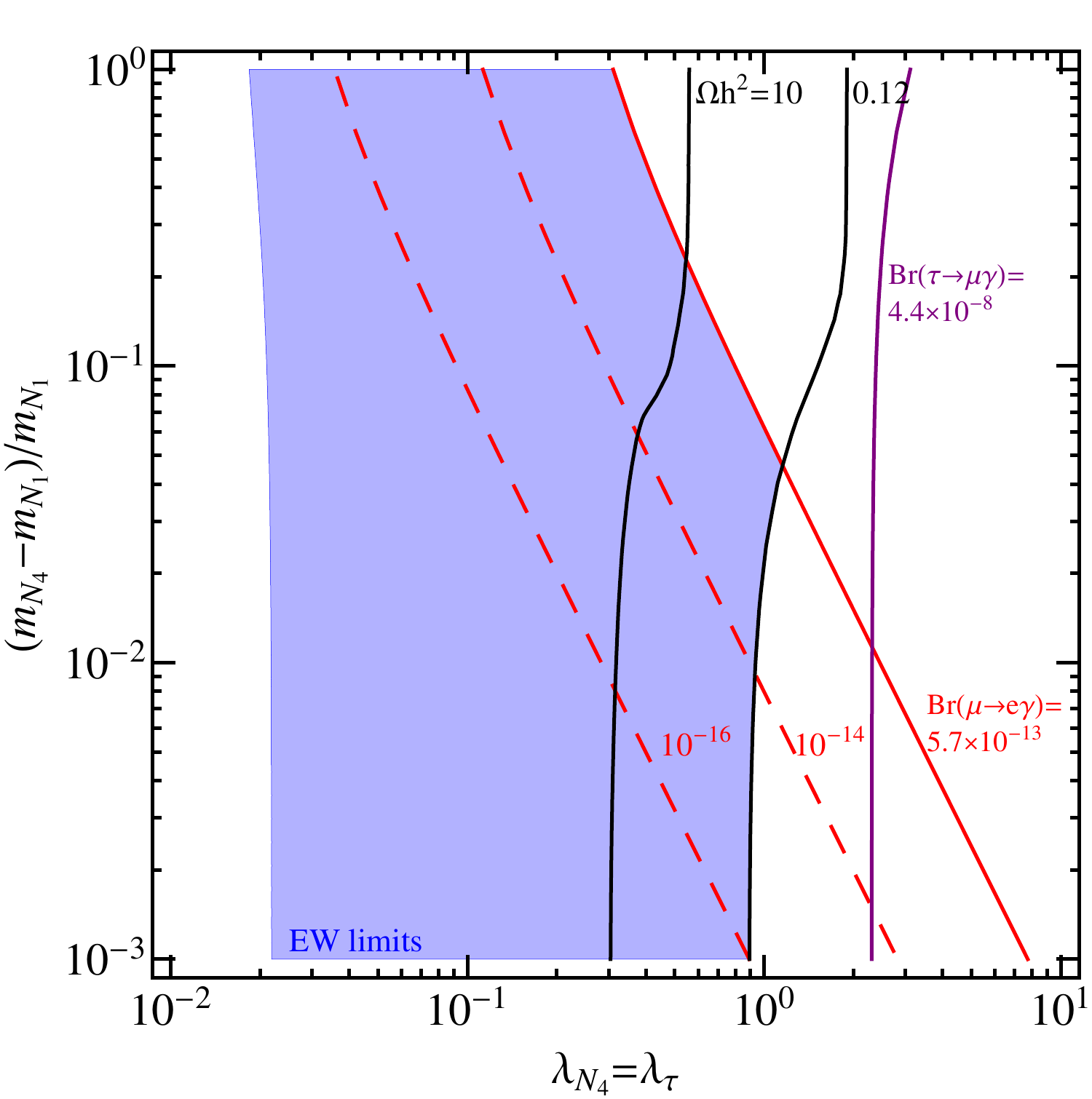} 
\includegraphics[width=0.45\textwidth]{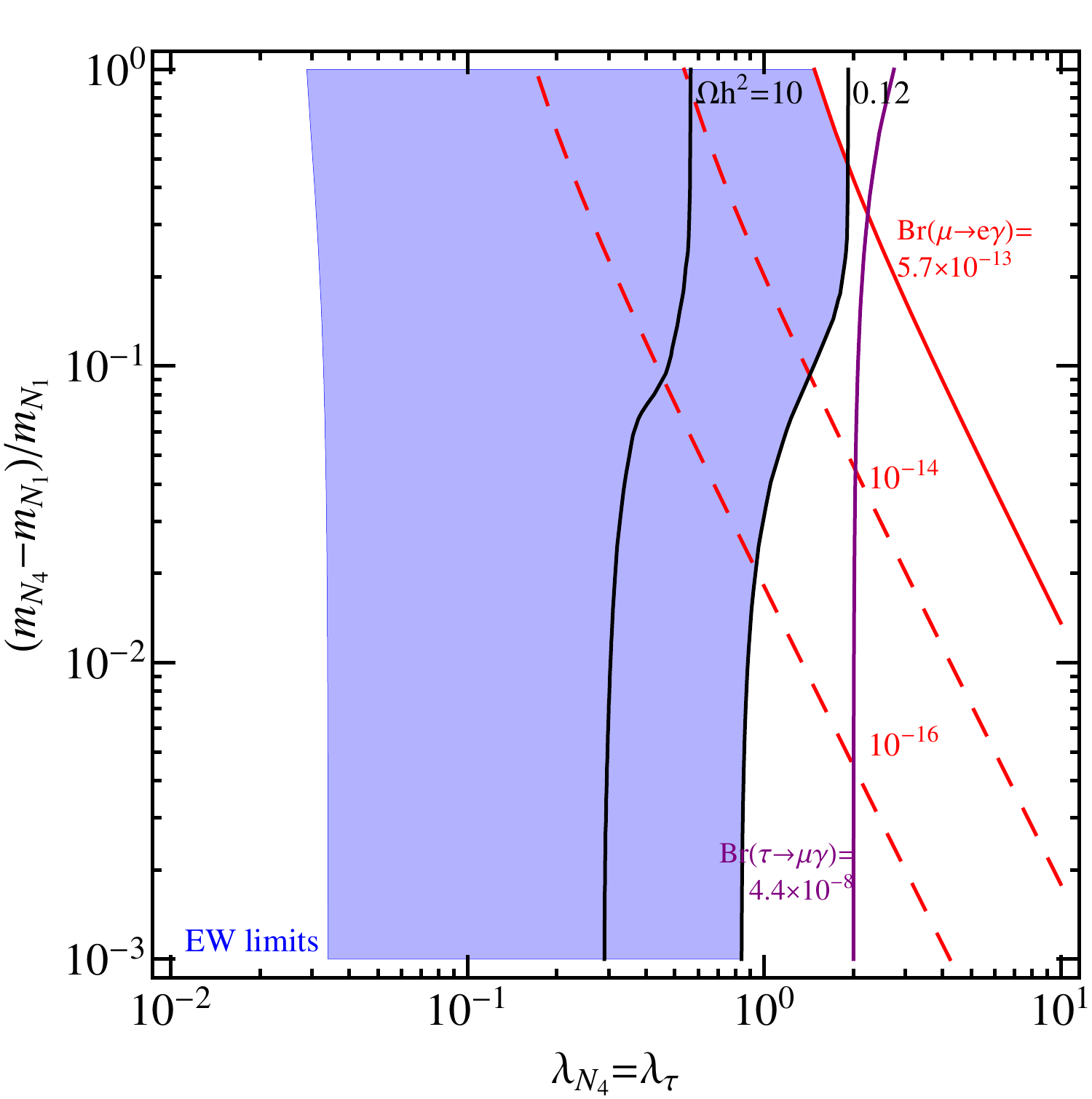}
\caption{Constraints on the parameter plane $\lambda = \lambda_{N_4} = \lambda_\tau$ and $\delta m_{41} = (m_{N_4}- m_{N_1})/m_{N_1}$ from $\mu\to e\gamma$ (red lines), $\tau\to \mu\gamma$ (purple lines), relic abundance $\Omega h^2$ (black curves) and EW precision measurements. The parameter space allowed by current experimental limits is shaded blue. The plots are for a NH light neutrino spectrum with a lightest neutrino mass of $m_\text{lightest}=10^{-2}$~eV~(left) and $m_\text{lightest}=10^{-1}$~eV~(right). The other model parameters are fixed as in the reference scenario of Tab.~\ref{tab:m_solve_yla}.}
\label{fig:r_la}
\end{figure}
Both the LFV observables $\text{Br}(\mu\to e\gamma)$,  $\text{Br}(\tau\to \mu\gamma)$ and the relic density $\Omega h^2$ depend delicately on the $\lambda$ couplings. In addition, the mass difference between the heavy neutrinos $N_1$ and $N_4$ suppresses $\text{Br}(\mu\to e\gamma)$ and $\Omega h^2$ through a GIM-like mechanism and its effect on the DM co-annihilation, respectively. This is shown in Fig.~\ref{fig:r_la}, where we plot the contours of these observables in the parameter plane of $\lambda_{N_4} = \lambda_\tau$ and $(m_{N_4} - m_{N_1})/m_{N_1}$. The other free parameters are set according to the reference scenario of Tab.~\ref{tab:m_solve_yla}. The two panels correspond to a NH light neutrino scenario with $m_\text{lightest}  = 0.01$~eV (left) and $m_\text{lightest}  = 0.1$~eV (right). The corresponding plots in the IH case would look very similar especially for $m_\text{lightest}  = 0.1$~eV as it corresponds to the quasi-degenerate regime. The red solid contour corresponds to the current experimental limit on $\text{Br}(\mu\to e\gamma)$ whereas the dashed red curves denote expected future sensitivities. The purple solid contour corresponds to the current experimental limit on $\text{Br}(\tau\to \mu\gamma)$, which unlike $\text{Br}(\tau\to \mu\gamma)$ is basically insensitive to the mass difference between $N_1$ and $N_4$.  
The solid black contours correspond to the relic abundances of $\Omega h^2 = 0.12$ and $\Omega h^2 = 10$ as denoted. As discussed in Section~\ref{sec:DM}, we do not explicitly take into account the annihilation through the $Z'$ in our model. We therefore interpret the experimental value $\Omega h^2 = 0.12$ as the lower limit for the abundance, with the understanding that the $Z'$ annihilation will further suppress the predicted value.

In addition to LFV and the DM abundance, the other important constraint on our model comes from the EW precision data, which limits the mixing between the active and sterile neutrinos and therefore the Yukawa couplings $y_{\alpha i}$. With the absence of standard Yukawa couplings of heavy neutrinos with the electron doublet, the relevant limits on the light-heavy mixing are \cite{delAguila:2008pw}
\begin{align}
	|V_{\mu i}|^2  &< 3 \times 10^{-3}, \nonumber\\
	|V_{\tau i}|^2 &< 6 \times 10^{-3},
\end{align}
with $i = 2, 3$. These limits can be translated to the Yukawa couplings using $V_{\alpha i} = y_{\alpha i} v / m_{N_i}$. In the given scenario with $m_{N_{2,3}} \approx 2$~TeV, they are therefore of the order $y_{\alpha i} < 10^{-1}$. As mentioned in the previous section, the experimental bounds on the LFV processes $\tau\to\mu\gamma$ and $\tau\to e\gamma$ put similar constraints on the Yukawa couplings, $y_{\alpha i} \lesssim 10^{-1}$, for heavy neutrino masses in the TeV region.

As the Yukawa couplings scale inversely with the $\lambda$ couplings as $y_{\alpha i} \approx 10^{-2}/\lambda^2$ in our scenario (see Fig.~\ref{fig:y_la_sol}~(left)), the EW limits will put a lower bound on $\lambda$ as shown in Fig.~\ref{fig:r_la}. The blue shaded regions denote the allowed parameter space when combining all the above constraints. Especially interesting in this model is the fact that the $\lambda$ couplings are bounded from above by the DM abundance and LFV, and from below by the EW precision data (and possibly $\tau$ LFV) through consistency with the neutrino data. The allowed region essentially covers the natural range $\lambda \approx 0.1 - 1$. Larger values for $\lambda$ are strictly disallowed as the resulting DM annihilation will be too strong. Unless the relative mass splitting between the heavy DM neutrino $N_1$ and the neutrino $N_4$ is larger than $\mathcal{O}(1)$, the current limit on $\text{Br}(\mu\to e\gamma)$ provides a weaker constraint than the relic abundance limit. However, future LFV searches will be able to probe the interesting region $\lambda \approx \mathcal{O}(1)$ and $\delta m_{41} = (m_{N_4} - m_{N_1})/m_{N_1} \approx 0.01 - 0.1$. Compared to the strong dependence $Br(\mu\to e\gamma) \propto \delta m_{41}^2$, the relic abundance is relatively weakly affected by $\delta m_{41}$, with a suppression of the abundance for $\delta m_{41} \lesssim 0.1$ due to enhanced co-annihilation. In addition, the current limit on $\text{Br}(\tau\to \mu\gamma)$ cannot give a further constraint since from Eq.~\eqref{eq:br_sim}, for $\la \sim 1$ with TeV scale $m_{\phi}$ and $m_{N}$, the predicted $\tau \to \mu \ga$ decay is smaller than the current bound.

\begin{figure}[t!]
\centering
\includegraphics[width=0.45\textwidth]{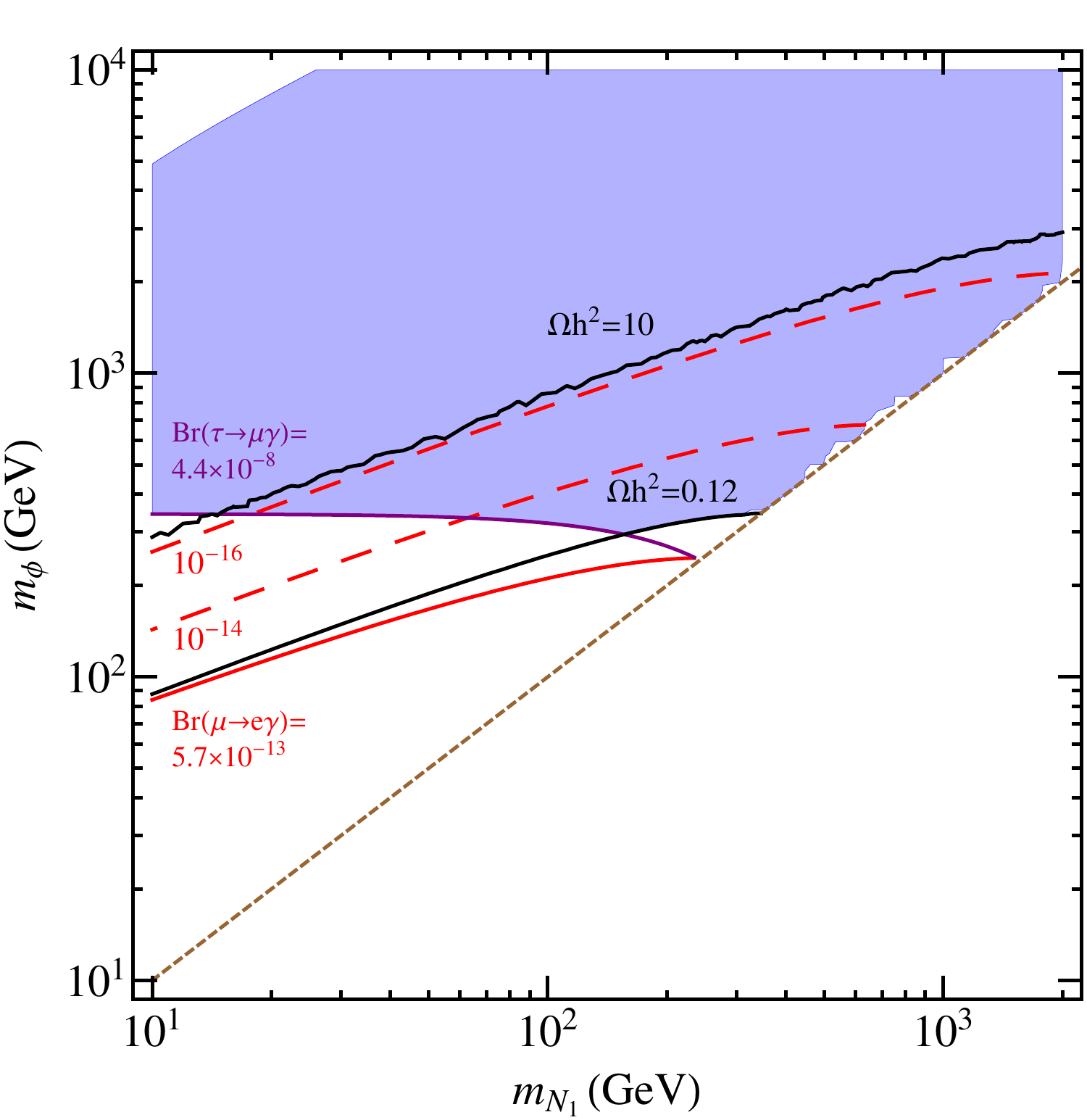}
\includegraphics[width=0.45\textwidth]{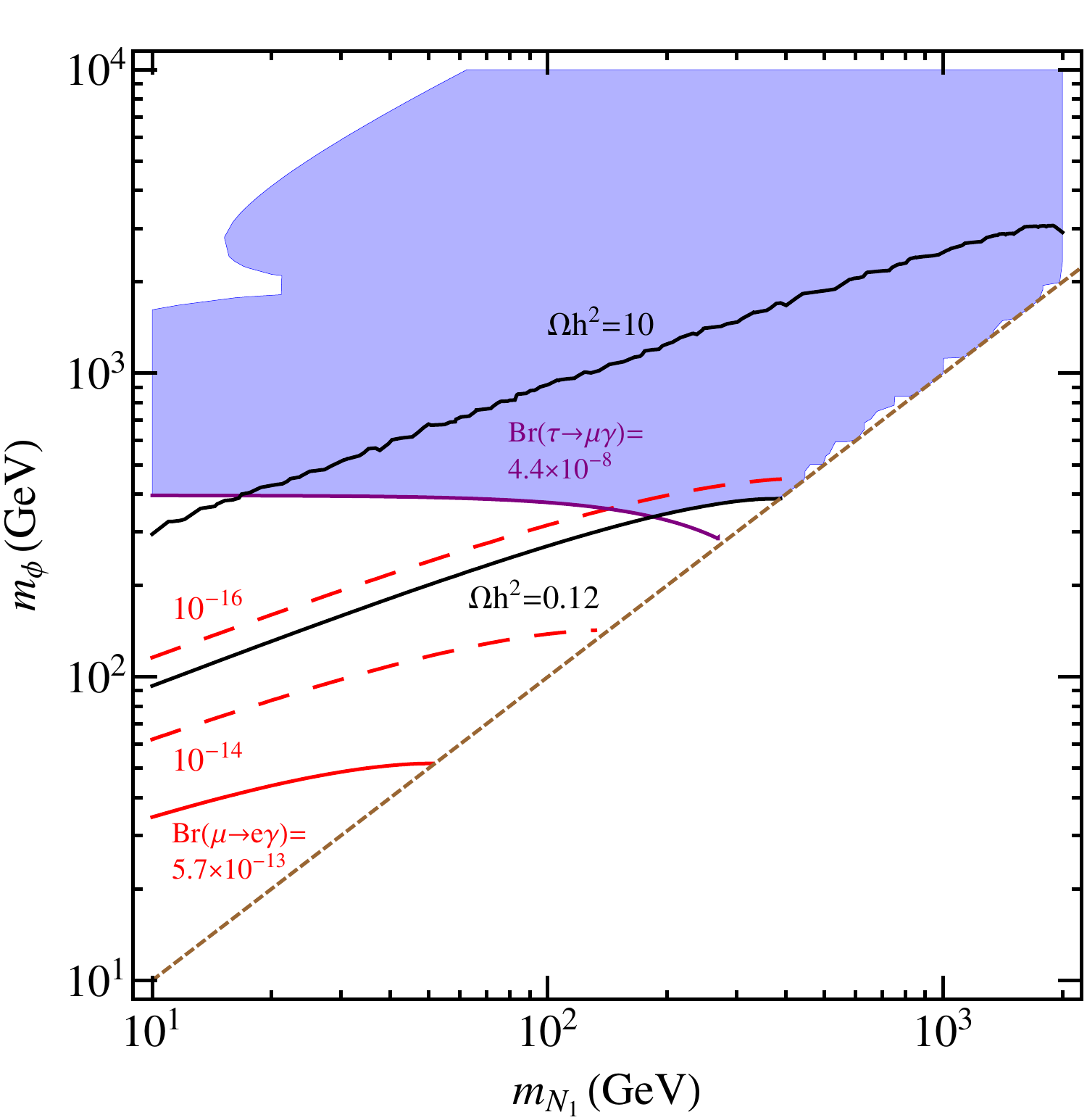}
\caption{As Fig.~\ref{fig:r_la} but in the parameter plane of $m_{N_1}$ and $m_\Phi$. The plots are for a NH light neutrino spectrum with a lightest neutrino mass of $m_\text{lightest}=10^{-2}$~eV~(left) and $m_\text{lightest}=10^{-1}$~eV~(right). The other model parameters are fixed as in the reference scenario of Tab.~\ref{tab:m_solve_yla}, but using a fixed relative mass splitting of $\delta m_{41} = 10^{-2}$.}
\label{fig:mn1_mphi}
\end{figure}
As demonstrated above, natural $\lambda$ coupling strengths $\lambda \sim \mathcal{O}(1)$ can nicely explain the observed relic abundance while still satisfying the bounds from EW precision data and LFV processes. In this case, the mass splitting between the lightest and second-lightest heavy neutrino should be roughly smaller than their absolute masses to also evade the LFV bound. Adopting the fixed values $\lambda = \lambda_{N_4} = \lambda_\tau = 0.5$ and $\delta m_{41} = 10^{-2}$, we now look at the dependence on the DM mass $m_{N_1}$ and the exotic Higgs mass $m_\Phi$. This is shown in Fig.~\ref{fig:mn1_mphi}, where we plot all observables in this parameter plane, analogously to Fig.~\ref{fig:r_la}. We do not consider the region below the diagonal as $N_1$ is not the DM particle in this case. Otherwise, the plot demonstrates that the mass of $N_1$ can range between 10~GeV and $10^3$~GeV (keep in mind that the relative splitting $\delta m_{41}$ is fixed to suppress LFV), whereas the Higgs mass $m_\Phi$ has to be $m_\Phi \gtrsim 100$~GeV. The plot also shows that both the LFV branching ratio $\text{Br}(\mu\to e\gamma)$ and the relic abundance $\Omega h^2$ behave similarly in the $m_\phi - m_N$ plane. In fact, for $m_\text{lightest} = 10^{-2}$~eV (left plot), the region probed by the current and future $\text{Br}(\mu\to e\gamma)$ sensitivities essentially coincides with the parameter space preferred by the observed relic abundance, $0.12 < \Omega h^2$. Moreover, the bound on $\text{Br}(\tau\to \mu\gamma)$ gives an additional constraint especially on the small $m_{N_1}$ and $m_{\phi}$ region. The corresponding plots in the IH light neutrino scenario are again very similar to the NH case shown here.

\begin{figure}[t]
\centering
\includegraphics[width=0.48\columnwidth]{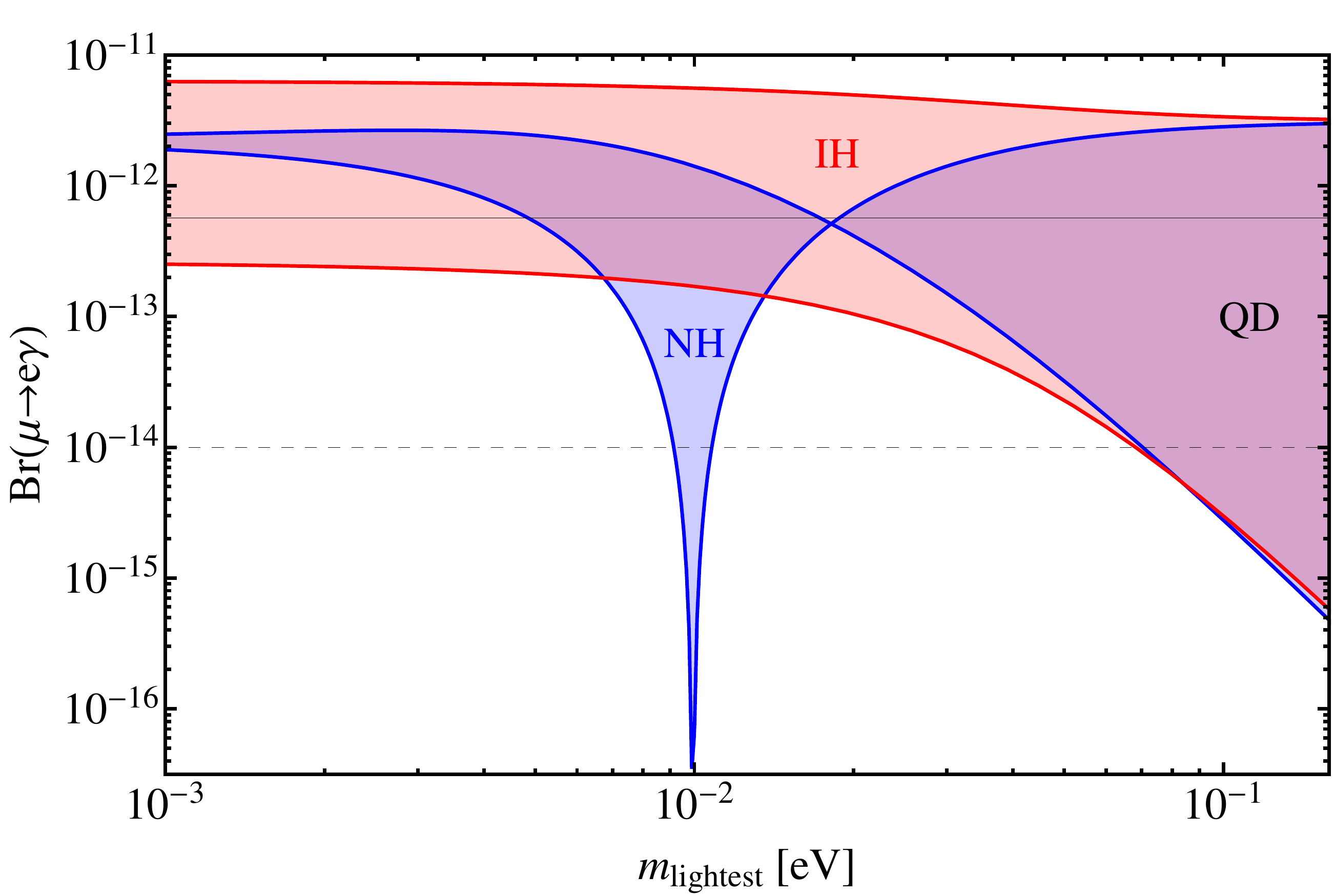} 
\includegraphics[width=0.48\columnwidth]{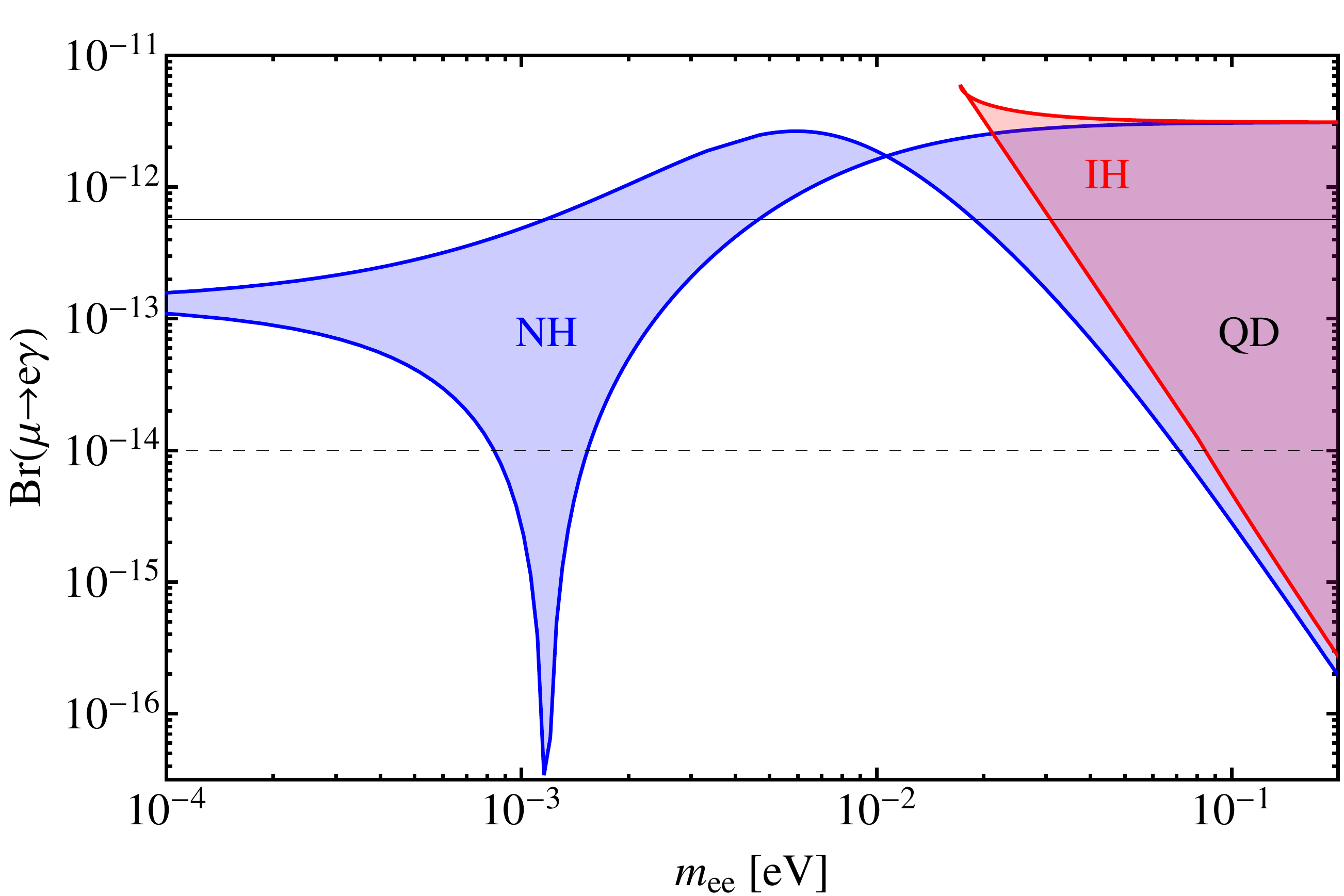}
\caption{LFV branching ratio $\text{Br}(\mu\to e\gamma)$ as a function of the lightest neutrino mass $m_{\text{lightest}}$ for NH and IH light neutrinos (left), and in correlation with the effective $0\nu\beta\beta$ mass $m_{ee}$ (right). The coloured bands correspond to the variation of the Majorana $CP$ phases $\alpha_1, \alpha_2$. The other model parameters are fixed as in the reference scenario of Tab.~\ref{tab:m_solve_yla}.}
\label{fig:muega_mnu1}
\end{figure}
So far we have essentially fixed the light neutrino mass matrix by using the specific values $m_\text{lightest} = 10^{-2}, 10^{-1}$~eV and adopting vanishing $CP$ phases $\delta = \alpha_1 = \alpha_2 = 0$. We now fix all the fundamental model parameters as shown Tab.~\ref{tab:m_solve_yla}, but look at the dependence on $m_\text{lightest}$, the light neutrino hierarchy and the Majorana $CP$ phases. For the latter we do not perform a full scan; instead, we adopt the extreme cases $\alpha_1 = \alpha_2 = 0$ and $\alpha_1 = \alpha_2 = \pi/2$, both corresponding to $CP$ conservation but with reverse $CP$ parities for the corresponding light neutrino states. $CP$-even observables like the LFV rate and the effective neutrinoless double beta decay mass $|m_{ee}|$ will in general range between the corresponding values. Particularly interesting feature is the dependence of $\text{Br}(\mu\to e\gamma)$ on the light neutrino parameters, and especially its correlation with the effective neutrinoless double beta decay mass. This is shown in Fig.~\ref{fig:muega_mnu1}~(left), where we plot $\text{Br}(\mu\to e\gamma)$ as a function of the lightest neutrino mass $m_\text{lightest}$ in the case of NH and IH neutrinos, using the denoted values for the Majorana $CP$ phases $\alpha_1$ and $\alpha_2$. The shaded areas denote the range in $\text{Br}(\mu\to e\gamma)$ for general Majorana $CP$ phases. The behaviour is actually quite similar to the corresponding plot for the effective $0\nu\beta\beta$ mass $|m_{ee}|$ as a function of $m_\text{lightest}$, with a dip around $m_\text{lightest} \approx 10^{-2}$~eV in the NH case with $\alpha_1 = \alpha_2 = \pi/2$. This is because in this regime the neutrino mass matrix element $(m^{\text{obs}}_\nu)_{12}$ is suppressed leading to a corresponding suppression of $\lambda_\mu$, cf. Section~\ref{fitting_nu_data}. The right plot in Fig.~\ref{fig:muega_mnu1} shows the same information but correlating $\text{Br}(\mu\to e\gamma)$ with the effective $0\nu\beta\beta$ mass $m_{ee}$. As the cancellations of contributions in $\text{Br}(\mu\to e\gamma)$ and $m_{ee}$ occur for different values of $m_\text{lightest}$, both observables provide complementary sensitivity. Specifically, the cancellation in $m_{ee} \to 0$ in the NH neutrino case corresponds to large $\text{Br}(\mu\to e\gamma)$.

\section{Conclusions}
\label{sec:conclusions}

In this work, we extend the scotogenic model proposed in Ref.~\cite{Ma:2006km} with an additional $U(1)_X$ gauge group under which heavy neutrinos and SM fermions are charged. Starting with three heavy neutrinos $N_{1,2,3}$, we assign an odd $U(1)_X$ charge to the lightest heavy neutrino $N_1$ and the additional $SU(2)_L$ scalar doublet $\phi$, while the other leptons carry even charges. The $U(1)_X$ gauge group is broken by the vacuum expectation values of scalars of even charges, leading to a residual $Z_2$ symmetry such that the lightest of the particles with odd charges is stable and the DM candidate. We assume that it is the lightest right-handed neutrino $N_1$ in this work. Moreover, an additional neutrino $N_4$ with the opposite $U(1)_X$ charge to $N_1$ is needed to make the model anomaly-free.   

The light neutrino masses are generated by both the seesaw type-I contribution and the radiative corrections with $U(1)_X$ odd-charge particles in loops. Meanwhile, the same interactions also induce LFV processes and contribute to DM annihilations. Therefore, we have an intriguing interplay among the neutrino observables, charged lepton flavour violation and the DM relic density. On the other hand, the seesaw type-I Yukawa couplings are forced to be relatively large~($10^{-2}\sim 10^{-3}$) in order to have sizeable loop contributions to the neutrino mass and reproduce the neutrino oscillation observables.

We have shown that the couplings $\la_i$ of the SM lepton doublets to $\phi$ and $N_{1,4}$ are naturally of $\mathcal{O}(1)$: Large $\la_i$ will lead to an insufficient DM density and large $\text{Br}(\tau (\mu) \to e \ga)$~(when the mass splitting $m_{N_4}-m_{N_1}$ is large), while small $\la_i$ require large Yukawa couplings, disfavoured by electroweak precision tests. In addition, the DM density and $\text{Br}(\mu \to e \ga)$ probe similar mass scales: The mass $m_\phi$ is bound from below, $m_{\phi} \gtrsim 100$ GeV, in order not to violate the DM density bound and the $\text{Br}(\tau(\mu) \to e \ga)$ constraint. Finally, in the NH case, the $\text{Br}(\mu \to e \ga)$ as a function of the lightest left-handed neutrino mass $m_\text{lightest}$ exhibits a dip around $m_\text{lightest}\sim 10^{-2}$ eV, similar to the behavior of the effective $0\nu\beta\beta$ mass but occurring for a different value of $m_\text{lightest}$.   

In this work we have focused on the neutrino mass generation mechanism, the charged lepton flavour violation and the DM relic density in this model. As an outlook, we would like to briefly comment on other possible consequences. Searches on the heavy right-handed neutrinos $N$ have been performed based on same-sign di-lepton events at the LHC~\cite{Aad:2011vj, Chatrchyan:2012fla, ATLAS:2012mn}. The constraints roughly apply for heavy neutrinos of the order of a few hundred GeV with certain assumptions on the production mechanism of $N$. In terms of the light-heavy neutrino mixing, the LHC bound is actually weaker than the EW precision test limit from Ref.~\cite{Atre:2009rg}. On the other hand, the relatively large Yukawa couplings in our model, required to produce sizable loop-induced neutrino masses, and also the additional $Z^\prime$ exchange production mechanism, increase the potential of discovering TeV scale $N$ at the LHC with $\sqrt{s}=13$ TeV. 

The most stringent bounds on $m_{Z^\prime}$ come from the di-lepton resonance searches~\cite{Chatrchyan:2012oaa, Aad:2014cka} with $m_{Z^\prime} \gtrsim 2.9$ TeV, assuming SM $Z$ couplings to fermions. A DM $N_1$ can be produced at the LHC via the $Z^\prime$ exchange. This process is constrained by null results on mono-jet and mono-boson searches at the LHC. See, for example, recent studies by CMS and ALTAS~\cite{Chatrchyan:2012me, Aad:2014vka}. The bounds become weak when the DM mass is above 1~TeV. We did not specify the gauge coupling $g_{Z^\prime}$ and it is not required to be large since the $\phi$ exchange processes alone can account for the correct DM density. In this case, the LHC DM search bounds will become irrelevant.

The addition of the $SU(2)_L$ doublet $\phi$ in our model can contribute to the electroweak $S$ and $T$ parameters \cite{Li:1992dt, Zhang:2006vt, Branco:2011iw}. As long as the mass splitting between the neutral and charged components is small, i.e. the custodial symmetry is approximate, $\phi$ contributions to $S$ and $T$ are under control. The charged component of $\phi$ will contribute to $H \to \gamma\gamma$~\cite{Arhrib:2012ia, Goudelis:2013uca}, depending on the strength $c_1$ of the interaction $(\tilde{H} \cdot H)(\phi^\dag \cdot\phi)$ and the mass of $\phi^{\pm}$. For $m_{\phi^\pm} \gtrsim 150$ GeV and $c_1 \approx 0.5$, the $H \to \ga\ga$ branching ratio could be very close to the SM prediction, consistent with the LHC data~\cite{ATLAS:2013oma, Khachatryan:2014ira}. 

The mechanism of neutrino mass generation and the nature of Dark Matter are two crucial issues so far unaccounted for. While certainly less than minimal, the hybrid model of neutrino mass generation discussed in this paper provides an example for the possibility to correlate the DM relic density, light neutrino properties and LFV rare decays. Furthermore, regions allowed by the experimental constraints correspond to naturally large couplings of the model which can be further probed by future experiments. Interestingly, and due to the required consistency in generating the observed neutrino parameters, searches for lepton flavour violation and measurements of electroweak precision observables can set both upper and lower limits on the Yukawa couplings of the model.

\section*{Acknowledgments}
The work of the authors was supported partly by the London Centre for Terauniverse Studies (LCTS), using funding from the European Research Council via the Advanced Investigator Grant 267352. The authors would like to thank Julia Harz for useful suggestions and a careful reading of the manuscript.

\appendix
\section{Anomaly Cancellation}
\label{sec:anomaly}
Here, we derive the anomaly constraints, including the SM gauge groups, $U(1)_X$ and gravity~($G$). Note that each lepton flavour is allowed to have a different $U(1)_X$ charge, while quarks may not have generation-dependent $U(1)_X$ charges due to flavour-changing neutral current processes for the $Z^\prime$ mass of interest~\cite{Carena:2004xs}. The relevant anomaly triangle diagrams are
\begin{itemize}
\item $[SU(3)]^2 U(1)_X$: $A_{33X}=3(2 X_Q  +X_U  + X_D)$,
\item $[SU(2)]^2 U(1)_X$: $A_{22X}= 9 X_Q + \sum_{j=1}^3  X_{L_j}  $,
\item $[U(1)_Y]^2 U(1)_X$: $A_{11X}= 2 X_Q +16 X_U  + 4 X_D + 2 \sum_{j=1}^3   ( X_{L_j} + 2 X_{E_j})$,
\item $U(1)_Y [U(1)_X]^2$: $A_{1XX}= 6(  X^2_Q -2 X^2_U + X^2_D ) - 2 \sum_{j=1}^3   (  X^2_{L_j} -X^2_{E_j})$,
\item $[U(1)_X]^3$: $A_{XXX}= 9 ( 2 X^3_Q + X^3_U + X^3_D ) +  \sum_{j=1}^3   ( 2 X^3_{L_j} + X^3_{E_j}) + \sum_{j=1}^n X^3_{N_j} $,
\item $[G]^2U(1)_X$: $A_{GGX}= 9 ( 2 X_Q + X_U + X_D ) +  \sum_{j=1}^3   ( 2 X_{L_j}  + X_{E_j}) + \sum_{j=1}^n X_{N_j} $,
\end{itemize}
where $Q$, $U$, $D$, $L$ and $E$ refer to the $SU(2)_L$ quark doublet, up-type singlet, down-type singlet, lepton doublet and singlet, respectively. $n$
is the number of heavy neutrinos in the model.\footnote{Again, we use the two component Weyl-spinor notation and all fields are left-handed. }
In addition, in our model, the Higgs boson is neutral under $U(1)_X$, which implies $X_{L_j}=-X_{E_j}$ in order to have three massive charged leptons
of different masses. Solving for $A_{33X}=0=A_{22X}=A_{11X}$, we have
\begin{align}
	X_Q=-X_U=-X_D= - \frac{1}{9} \sum_{j=1}^{3} X_{L_j},
\end{align}
and the remaining equations read
\begin{gather}
	\sum^3_{j=1} X^3_{L_j} + \sum^n_{j=1} X^3_{N_j}  = 0, \quad
	\sum^3_{j=1} X_{L_j}   + \sum^n_{j=1} X_{N_j}  = 0,
\label{eq:anomaly_can}
\end{gather}
where the second equation determines $X_{Q,U,D}$, and the first one puts constrains on $X_{N_j}$. In order to make the DM candidate $N_1$ stable,
we assume $X_{N_1}=-1$ and the other fermions have even charges as discussed in Section~\ref{sec:model}. In addition, the radiative neutrino mass matrix in Eq.~\eqref{eq:m_L} renders only one of three light neutrinos massive; therefore, at least one of $N_2$ and $N_3$ has to couple to the lepton doublet $L$ via Yukawa couplings. In summary, we start with $X_{N_1}=-1$ and $X_{N_2}=-2=- X_{L_2}$, where we have chosen $N_2$ to couple to $L_2$ without loss of generality. Given three heavy neutrinos $N_{(1,2,3)}$, one always ends up with $L_1=1$ or $L_3=1$. It allows a term like $L_1 H N_1$, leaving $N_1$ unstable.\footnote{Note that this conclusion does not depend on the assumption of integer charge. For $n=3$, Eq.~\eqref{eq:anomaly_can}  dictates only two possible solutions: $X_{L_1}=-X_{N_1}$ and $X_{L_3}=-X_{N_3}$ or $X_{L_1}=-X_{N_3}$ and $X_{L_3}=-X_{N_1}$.} With the minimum extension of the existing framework, one can include one additional heavy neutrino $N_4$, which has $X_{N_4}= - X_{N_1}$ to cancel the $N_1$'s anomaly contribution. As a result, there are three possibilities, where we set $X_{N_3}=X_{N_2}$ and restrain ourself to solutions of integer charge only for simplicity:
\begin{itemize}
\item $X_{N_4}= - X_{N_1}=1$, $X_{N_2}=X_{N_3}=-2$, $X_{L_\mu}=X_{L_\tau}=2$ and $X_{L_e}=0$,
\item $X_{N_4}= - X_{N_1}=1$, $X_{N_2}=X_{N_3}=-2$, $X_{L_e}=X_{L_\tau}=2$ and $X_{L_\mu}=0$,
\item $X_{N_4}= - X_{N_1}=1$, $X_{N_2}=X_{N_3}=-2$, $X_{L_e}=X_{L_\mu}=2$ and $X_{L_\tau}=0$,
\end{itemize}
denoted by $\eca$, $\mca$ and $\tca$, respectively.
In other words, we introduce $N_4$ to cancel $N_1$'s contributions at the cost of one of $X_{L}$'s being zero.

\section{DM Annihilation Cross Section}
\label{sec:sigmav}
We here collect the annihilation cross section $\sigma v$ for processes involving the $\phi$-exchange in the context of the $\eca$ case.
According to initial states, the results are divided into three pieces: $N^m_1 N^m_1$, $N^m_4 N^m_4$ and $N^m_1 N^m_4$.
All of them consist of three contributions: $(\sigma v)_{tt}$, $(\sigma v)_{uu}$, $(\sigma v)_{tu}$, where $tt$~($uu$) denotes the square
of the $t$-~($u$-) channel amplitude and $tu$ refers to the interference term.

The basic building blocks are differential cross sections corresponding to $tt$, $uu$ and $tu$ as follows.
\begin{align}
	\frac{d(\sig v)_{tt}}{d c_\th}  &=   
	\frac{ m^2_N }{32 \pi}\frac{  \lee 2 - c_\th v   \rii^2   \lee 4- v^2 \rii   }
	{\lee    m^2_{\phi} \lee 4 - v^2   \rii   + m^2_{N}  \lee 4 -4 c_\th v + v^2 \rii       	
	\rii^2  }, \\
	\frac{d(\sig v)_{uu}}{d c_\th} &=   
	\frac{ m^2_N }{32 \pi}\frac{  \lee 2 + c_\th v   \rii^2   \lee 4- v^2 \rii   }
	{  \lee    m^2_{\phi} \lee 4 - v^2   \rii   + m^2_{N}  \lee 4 +4 c_\th v + v^2 \rii       
	\rii^2  },  \\
	\frac{d(\sig v)_{tu}}{d c_\th}  &=  
	- \frac{ m^2_N }{16 \pi}
	\frac{  \lee   4 - v^2   \rii^2     }
	{  \lee   16 \lee m^2_\phi + m^2_N   \rii^2   
	-8 v^2  \lee  m^4_\phi - m^4_N \lee 1 -2 c^2_\th \rii   \rii  + 
  v^4 \lee m^2_\phi  - m^2_N   \rii^2    \rii  }, 
\end{align}
where $c_\th\equiv \cos\th$ with $\th$ being the angle between the incoming and outgoing particle.
$m_N$ is the incoming particle mass. The total cross sections can be obtained by integrating over $c_\th$,
\begin{align}
	(\sigma v)_X = \int_{-1}^{1} d c_\th \frac{d(\sig v)_X}{d c_\th}.
\end{align}
Since we are mainly interested in the degenerate region of $m_{N_1}\sim m_{N_4}$, in order to simplify the computation, we use $m_N=\lee m_{N_1} + m_{N_4}\rii/2$ for co-annihilation processes which involve two different incoming particles. The $N_1 N_1$ cross section is
\begin{align}
	(\sig v)_{11} &= 
	\left(
		|\la_{N_4} \la_{N_4}^* U_{12} U_{12}^*|^2 +
			\sum_{\al=\mu,\tau} (|\la_\al \la_\al^* U_{22} U_{22}^*|^2 
												 + |\la_\al \la_{N_4}^* U_{22} U_{12}^*|^2 
												 + |\la_{N_4} \la_{\al}^* U_{12} U_{22}^*|^2) 
	\right) \nn\\
	&\times ((\sigma v)_{tt} + (\sigma v)_{uu} + (\sigma v)_{tu}), 
\end{align}
where $U_{ij}$ is the $(i,j)$ element of $U_{41}$ defined in Eq.~\eqref{eq:N1-N4_mixing}. The $N_4 N_4$ cross section is
\begin{align}
	(\sig v)_{44} &= \lee 
			|\la_{N_4} \la_{N_4}^* U_{11} U_{11}^*|^2 + 
			\sum_{\al=\mu,\tau} (|\la_\al \la_\al^* U_{21} U_{21}^*|^2   
		+ |\la_\al \la_{N_4}^* U_{21} U_{11}^*|^2 
		+ |\la_{N_4} \la_{\al} U_{11}  U_{21}|^2)  \rii \times \nn\\
&\times ((\sigma v)_{tt} + (\sigma v)_{uu} + (\sigma v)_{tu}).
\end{align}
The co-annihilation $N_1 N_4$ cross section is more complicated since the $uu$, $tt$ and $tu$ contributions have different coefficients that cannot be factored out as before. The cross sections for the different final states $L^\dag_\alpha L_\alpha$, $L^\dag_e L_\alpha$, $L^\dag_\alpha L_e$ and $L^\dag_e L_e$ ($\alpha = \mu, \tau$) are respectively given by
\begin{align}
	(\sig v)_{41}^{\al\al} &= 
	( |\la_\mu|^4 + |\la_\tau|^4) 
	\left( |U_{22} U_{21}^*|^2 (\sig v)_{tt}
			 + |U_{21} U_{22}^*|^2 (\sig v)_{uu}  
			 + \text{Re}[(U_{22} U^*_{21})^2] (\sig v)_{tu} 
	\right), \\
	(\sig v)_{41}^{ee} &=  
	|\la_{N_4}|^4 
	\left( |U_{12} U_{11}^*|^2 (\sig v)_{tt}
			 + |U_{11} U_{12}^*|^2 (\sig v)_{uu} 
			 + \text{Re}[(U_{12} U^*_{11})^2] (\sig v)_{tu}
	\right), \\
	(\sig v)_{41}^{\al e} &= (\sig v)_{41}^{e \al} \nn\\
	&=
	|\la_{N_4}|^2 (|\la_\mu|^2 + |\la_\tau|^2) 
	\left( |U_{22} U_{11}^*|^2 (\sig v)_{tt}
			 + |U_{21} U_{12}^*|^2 (\sig v)_{uu}  
			 + \text{Re}[U_{22} U^*_{11} U_{12} U^*_{21}] (\sig v)_{tu}
	\right).	
\end{align}

\bibliography{DM_zp}
\bibliographystyle{h-physrev}

\end{document}